\documentclass[acmsmall]{acmart}

\AtBeginDocument{%
  }

\acmJournal{TOIS}
\acmYear{2025} \acmVolume{1} \acmNumber{1} \acmArticle{1} \acmMonth{1}\acmDOI{10.1145/3722553}

\usepackage{booktabs}       % professional-quality tables
\usepackage{amsfonts}       % blackboard math symbols
\usepackage{xcolor}         % colors

\usepackage{tikz}
\usepackage{pgfplots}
\usepgfplotslibrary{statistics}
\usepackage{enumitem}

\usepackage{amsmath,amsthm,bm} % Math packages
\usepackage{cancel} 
\usepackage{subcaption}
\usepackage{graphicx}

\usepackage{array}
\newcolumntype{P}[1]{>{\centering\arraybackslash}p{#1}}

\usepackage{wrapfig} %文字环绕

\usepackage{hyperref}
\usepackage{url}

\usepackage{algorithm}
\usepackage{algpseudocode}
\newtheorem{theorem}{Theorem}

\newtheorem{definition}{Definition}

\def\Uin{\bm{o}_{in}^{u}} %user intrinsic representation
\def\Uex{\bm{o}_{ex}^{u}} %user extrinsic representation

\def\Iin{\bm{o}_{in}^{v}} %Item intrinsic representation
\def\Iex{\bm{o}_{ex}^{v}} %Item extrinsic representation

\def\Gin{\bm{o}_{in}} %General intrinsic representation
\def\Gex{\bm{o}_{ex}} %General extrinsic representation

\def\urep{\bm{u}} %user profile representation
\def\irep{\bm{v}} %item profile representation
\def\crep{\bm{c}} %context state representation

\def\I{\mathcal{I}} %mutual information symbol

\def\sim{\text{sim}}

\def\ufrep{\bm{z}^{u}} %user feature representation
\def\ufie{f^{u}_{ie}} %user f_ie function
\def\ifie{f^{v}_{ie}} %item f_ie function
\def\mR{\mathbb{R}} % real value space

\newcounter{relctr} %% <- counter for relations
\everydisplay\expandafter{\the\everydisplay\setcounter{relctr}{0}} %% <- reset every eq
\newcommand\labelrel[2]{%
  \begingroup
    \refstepcounter{relctr}%
    \tiny
    \stackrel{\textnormal{(\alph{relctr})}}{\mathstrut{#1}}%
    \originallabel{#2}%
  \endgroup
}
\AtBeginDocument{\let\originallabel\label} %% <- store original definition

\definecolor{mycolor1}{RGB}{246, 126, 0}
\definecolor{mycolor2}{RGB}{120, 81, 151}
\definecolor{mycolor3}{RGB}{84, 144, 196}
\definecolor{mycolor4}{RGB}{205, 35, 33}

\pgfplotsset{linestyle/.style={%
        font=\rmfamily\Labelsize,
        width=0.27\textwidth,
        height=4.5cm,
        mark size=1.1pt,
        ylabel near ticks,
        xlabel near ticks,
        xtick=data,
        label style = {font=\small},
        tick label style = {font=\small, yshift=0.5ex},
        ylabel shift = -6 pt, 
        xlabel shift = -5 pt,
        title style={yshift=-1.2ex,font=\small},
        legend image post style={scale=0.5},
        every axis plot/.append style={ultra thick},
        legend style={font=\small, mark size=4pt},
        legend columns=1, legend style={/tikz/column 2/.style={column sep=0.5pt}},
        mark options={scale=1}}}

\pgfplotsset{barstyle/.style={%
    ybar,
    ylabel near ticks,
    xlabel near ticks,
    height=4.5cm,
    width=0.30\textwidth,
    label style={font=\small},
    tick label style={font=\small, yshift=0.6ex},
    title style={yshift=-1.2ex,font=\small},
    ylabel shift = -5 pt, 
    xlabel shift = -2 pt,
    tickwidth         = 3pt,
    xtick=data,
    legend style={font=\small, mark size=1pt},
    legend columns=2, legend style={/tikz/column 1/.style={column sep=0.1pt}},
}}

\pgfplotsset{boxstyle/.style={
    boxplot/draw direction=y,
    ylabel near ticks,
    xlabel near ticks,
    height=4.5cm,
    width=0.30\textwidth,
    cycle list={{mycolor1},{mycolor3}},
    label style={font=\small},
    tick label style={font=\small, yshift=0.6ex},
    xtick={1,2},
    ticklabel style={yshift=-0.5ex},
    title style={yshift=-1.2ex,font=\small},
    ylabel shift = -5 pt, 
    xlabel shift = -2 pt,
    legend style={font=\small, mark size=1pt},
    legend columns=1, legend style={/tikz/column 1/.style={column sep=0.1pt}},
}}

\pgfplotsset{xbarstyle/.style={%
    xbar,
    width=0.225\textwidth,
    label style={font=\small},
    tick label style={font=\small},
    axis x line*=bottom,
    axis y line*=none,
    height=4.5cm,
    title style={yshift=-0.5ex,font=\small},
    ylabel shift = -5 pt, 
    xlabel shift = -5 pt,
    ylabel near ticks,
    xlabel near ticks,
    tickwidth         = 3pt,
    ytick=data,
    bar width=1mm,
    xlabel style={font=\small}
}}

\begin{document}

\title{Intrinsic and Extrinsic Factor Disentanglement for Recommendation in Various Context Scenarios}

\author{Yixin Su}
\authornote{Both authors contributed equally to this research.}
\authornote{Yixin Su did this work when he was an intern at Alibaba Group.}
\email{yixin.su@outlook.com}
\affiliation{%
\institution{School of Computer Science and Technology, Huazhong University of Science and Technology}
\city{Wuhan}
\country{China}
}

\author{Wei Jiang}
\email{wwjiangwei@hotmail.com}
\authornotemark[1]
\affiliation{%
\institution{Alibaba Group}
\city{Hangzhou}
\country{China}
}

\author{Fangquan Lin}
\email{fangquan.linfq@alibaba-inc.com}
\affiliation{%
\institution{Alibaba Group}
\city{Hangzhou}
\country{China}
}

\author{Cheng Yang}
\email{charis.yangc@alibaba-inc.com}
\affiliation{%
\institution{Alibaba Group}
\city{Hangzhou}
\country{China}
}

\author{Sarah M. Erfani}
\email{sarah.erfani@unimelb.edu.au}
\affiliation{%
\institution{The University of Melbourne}
\city{Melbourne}
\country{Australia}
}

\author{Junhao Gan}
\email{junhao.gan@unimelb.edu.au}
\affiliation{%
\institution{The University of Melbourne}
\city{Melbourne}
\country{Australia}
}

\author{Yunxiang Zhao}
\email{zhaoyx1993@163.com}
\authornote{Corresponding authors.}
\affiliation{%
\institution{Laboratory of Advanced Biotechnology, Beijing Institute of Biotechnology}
\city{Beijing}
\country{China}
}

\author{Ruixuan Li}
\email{rxli@hust.edu.cn}
\affiliation{%
\institution{School of Computer Science and Technology, Huazhong University of Science and Technology}
\city{Wuhan}
\country{China}
}

\author{Rui Zhang}
\email{rayteam@yeah.net}
\authornotemark[3]
\affiliation{%
\institution{School of Computer Science and Technology, Huazhong University of Science and Technology (www.ruizhang.info)}
\city{Wuhan}
\country{China}
}

\renewcommand{\shortauthors}{Su, et al.}

\begin{abstract}

In recommender systems, the patterns of user behaviors (e.g., purchase, click) may vary greatly in different contexts (e.g., time and location).
This is because user behavior is jointly determined by two types of factors: \textit{intrinsic factors}, which reflect consistent user preference, and \textit{extrinsic factors}, which reflect external incentives that may vary in different contexts. 
%Recent studies involved intrinsic and extrinsic factor learning to improve recommendation accuracy. For example, some studies leverage time context to disentangle users' long-term interests as intrinsic factors and short-term interests as extrinsic factors. However, a user's extrinsic factors may be influenced by various contexts simultaneously, while these studies only rely on a single, pre-defined context (e.g., time) to learn the factors. Therefore, these studies lack the ability to capture the interplay of various contexts and their joint influence, which is crucial for effective factor learning. 
%Differentiating intrinsic factors from extrinsic ones helps learn user behaviors better, but existing studies have only considered differentiating them in only one context (e.g., time or location). They cannot differentiate intrinsic from extrinsic factors in various contexts such as time, location, weather, and social settings at the same time.
Differentiating between intrinsic and extrinsic factors helps learn user behaviors better. However, existing studies have only considered differentiating them from a single, pre-defined context (e.g., time or location), ignoring the fact that a user's extrinsic factors may be influenced by the interplay of various contexts at the same time. 
%As a result, they may fail to accurately differentiate intrinsic and extrinsic factors.
%In this paper, we propose IEDR, a generic framework that learns disentangled intrinsic and extrinsic factors from various contexts at the same time, enabling more accurate differentiating of factors and hence the improvement of recommendation accuracy.
In this paper, we propose the Intrinsic-Extrinsic Disentangled Recommendation (IEDR) model, a generic framework that differentiates intrinsic from extrinsic factors considering various contexts simultaneously, enabling more accurate differentiation of factors and hence the improvement of recommendation accuracy.
IEDR contains a context-invariant contrastive learning component to capture intrinsic factors, and a disentanglement component to extract extrinsic factors under the interplay of various contexts. The two components work together to achieve effective factor learning.
Extensive experiments on real-world datasets demonstrate IEDR's effectiveness in learning disentangled factors and significantly improving recommendation accuracy by up to 4\% in NDCG.
\end{abstract}






\maketitle

\section{Introduction}
Recommender systems \cite{liu2024multimodal,shen2024pmg,wang2023missrec,zhu2022bars} aim to predict the probability of a user's behavior (e.g., purchase, click) on a given item.
This is a challenging task since a user's behavior may vary significantly across different \textit{contexts} (e.g., time, location, and social setting).
For example, considering the context of social settings (e.g., alone vs. with friends), when recommending food, a user may prefer healthy food like steamed vegetables and salad when being alone, but may prefer more diverse food suitable for sharing like hot pot or pizza when gathering with friends.
%This variation in user behaviors across different contexts highlights their complex nature.
This context-dependent variation in user behaviors underscores their complex nature.
Psychological research has devoted great efforts to understanding this phenomenon, and reveals that user behaviors are influenced by two types of factors: \textit{intrinsic} and \textit{extrinsic} factors \cite{benabou2003intrinsic,vallerand1997toward}, distinguished by whether they can be influenced by context changes.
An intrinsic factor, which is often stable for a user across different contexts, is an internal motivation for inherent satisfaction. 
In our food recommendation example, the preference for healthy food when eating alone could be driven by intrinsic factors such as personal health goals or taste preferences.
In contrast, an extrinsic factor, which is an external motivation stimulated by the contexts, often varies when contexts change \cite{ryan2000intrinsic}.
The choice of more diverse food when gathering with friends could be influenced by extrinsic factors such as the social setting.
%To illustrate, still in a food recommendation scenario, a user's behavior to frequently choose hot soup may be motivated by intrinsic factors, such as a long-term health consciousness. On the other hand, the same user's choice of having a hot pot may be influenced by extrinsic factors, such as the social setting of gathering with friends.
Therefore, to better understand user behaviors and provide more accurate recommendations, it is crucial yet challenging for recommender systems to effectively capture and differentiate between intrinsic and extrinsic factors in various contexts.
Existing studies that aim to differentiate between intrinsic and extrinsic factors consider only a single, pre-defined context, e.g., time \cite{yu2019adaptive,duan2023long} or location \cite{hidasi2016session,li2017learning}.
%However, these methods are limited since they rely on a single, pre-defined context (e.g., time or location) to capture intrinsic and extrinsic factors. This oversimplifies the complex nature of user behavior, which is often influenced by the interplay of various contexts simultaneously.
However, in reality, user behaviors are often influenced by the interplay of various contexts simultaneously. These methods may not be able to accurately capture user behaviors
%capture users' preferences 
, especially when contexts change (an example will be given in the next paragraph).
%Moreover, these methods are designed specifically for the context they considered. For example, \citet{li2017learning} leveraged location context to differentiate intrinsic and extrinsic factors by assuming the choice of a long geographical distance place is more influenced by intrinsic factors and vice versa. 
Moreover, these methods are designed specifically for the pre-defined context. For example, \citet{li2017learning} leverages location context to differentiate intrinsic and extrinsic factors. They incorporate a context-specific assumption into their model that the choice of a long geographical distance place is more influenced by intrinsic factors and vice versa.
%Therefore, it is hard for these methods to extend to other contexts in various contexts scenarios, e.g., this location-specific assumption cannot be adapted to social setting context. 
Consequently, it is difficult to extend these methods to scenarios where multiple types of contexts may affect the result. For instance, this location-specific assumption cannot be adapted to a social setting context.

\begin{figure*}[t]
\centerline{\includegraphics[width=0.9\textwidth]{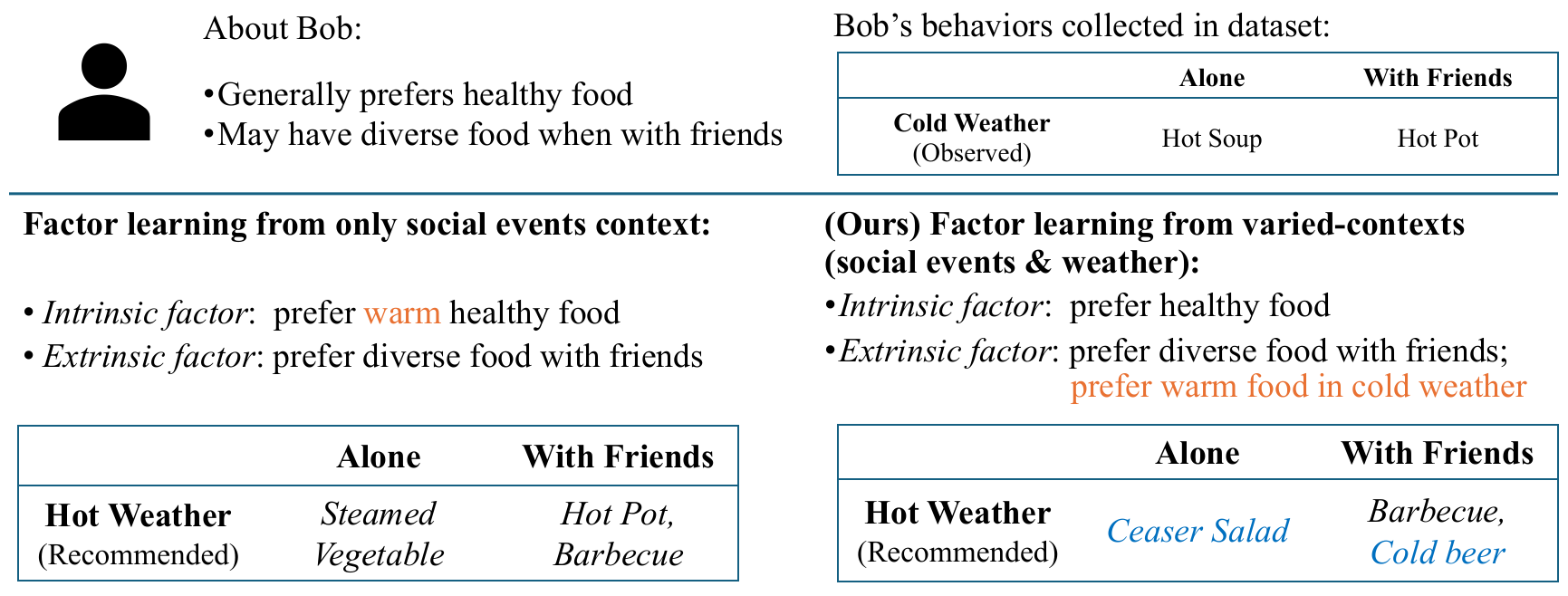}}
%\vskip -0.08in
\caption{
%An illustration of the difference between existing studies and our work in learning user preferences through an example of user's decision on taking an Uber. Existing studies consider only one specific context (e.g., time), while our work considers all relevant contexts, enabling better learning of intrinsic and extrinsic factors for more accurate recommendations, such as better capturing user's extrinsic influence of a promotional discount.
An example to compare existing work (consider only the context of social settings) and our approach (consider various contexts) in learning intrinsic and extrinsic factors. The upper part shows the preference fact (upper left) and observed behaviors (upper right) of a user Bob. The bottom part shows the possible factor learning results and corresponding recommendations of existing work (bottom left) and our approach (bottom right).
%Existing studies (upper right) only consider the time context (where the long and short arrows represent the user's long-short term preferences), leading to consistent extrinsic factors but failing to capture the influence of promotional discounts. In contrast, our work (lower right) considers all relevant contexts, such as time and discount, enabling better learning of extrinsic factors.
}
\label{fig:example}
\end{figure*} 

Given these limitations, in this paper, we aim to capture and differentiate between intrinsic and extrinsic factors from various contexts, thereby enhancing the ability to learn user behaviors.
To this end, we adopt an approach from a more fundamental perspective %focusing on whether the factors vary when contexts change, 
without introducing any context-specific assumptions.
%To this end, we consider to differentiate the two types of factors from a more fundamental perspective, focusing on whether they vary when contexts change, without introducing any context-specific inductive bias.
Under this general context condition, we first define intrinsic and extrinsic factors by focusing on whether these factors vary when contexts change.
%Focusing on whether the factors vary when contexts change, we first define intrinsic and extrinsic factors in a general context condition, i.e., we do not assume the number and types of contexts.
Following this definition, we propose an \underline{I}ntrinsic-\underline{E}xtrinsic \underline{D}isentangled \underline{R}ecommendation (IEDR) model, a general framework that can effectively capture the interplay of various contexts and differentiate intrinsic and extrinsic factors within them.
%capture and differentiate intrinsic and extrinsic factors under the interplay of various contexts. 
%Our model does not introduce any context-specific inductive bias, and hence achieved a unified learning of contexts for accurate intrinsic and extrinsic differentiation in various contexts scenarios.
%Figure \ref{fig:example} illustrates the difference between existing studies and our work in learning user preferences, using an example of a user's decision on taking an Uber. Existing studies consider only a single context (e.g., time), failing to capture the influence of extrinsic factors present in other contexts, such as promotional discounts. In contrast, our work considers diverse contexts, enabling more accurate learning of intrinsic and extrinsic factors, thus providing more accurate recommendations that account for the complex interactions among various contexts.
To illustrate the importance of accurately differentiating between intrinsic and extrinsic factors in scenarios with various contexts, consider the example in Figure \ref{fig:example}.
A user called Bob generally prefers healthy food but enjoys diverse food when gathering with friends (top left of the figure). The dataset happens to only contain Bob's behaviors in cold weather (top right of the figure), where Bob has steamed vegetables (warm healthy food) when alone and hot pot (diverse option) with friends.
Existing models differentiate between intrinsic and extrinsic factors from only one of the contexts, such as social settings (i.e., alone vs. with friends) in this example. They might incorrectly identify warm food preference as Bob's intrinsic factor (lower left of the figure).
This is because the model treats the weather context (i.e., cold vs. hot) as a regular feature rather than a context used for factor differentiation. The weather-dependent influence may show similar patterns across different social settings (e.g., warm foods are chosen either when alone or with friends), leading to weather-dependent extrinsic factors being mistakenly identified as intrinsic factors.
%In contrast, our model considers various contexts for factor learning (lower right of the figure), and accurately captures Bob's extrinsic factors like weather-dependent warmth preference in cold weather. 
In contrast, our model considers various contexts for differentiating the factors (lower right of the figure). Since a strong correlation may exist between weather and warm/cold food choices (e.g., most users may choose warm food in cold weather and cold food in hot weather), our model captures such weather-dependent preferences as extrinsic factors. Bob's choices of warm food all occur in cold weather, fitting well with the weather-dependent preference pattern (i.e., preferring warm food in cold weather). Therefore, our model can accurately capture such choices as being influenced by extrinsic factors.
When in hot weather scenarios (shown in the bottom two tables of the figure), existing models (left table) may incorrectly recommend hot food due to misidentified intrinsic factors. In comparison, our model (right table) adapts to the weather context, recommending more suitable cold options like Caesar salad and cold beer.
The IEDR framework consists of two main modules: a recommendation prediction (RP) module and a contrastive intrinsic-extrinsic disentangling (CIED) module.
To better capture the interplay among different contexts, the RP module constructs various contexts into a graph structure, where each context is represented as a node and their interplay (interactions) is represented as edges, and a complete graph is constructed. By applying graph learning algorithms to this context graph, the model can comprehensively learn the complex relationships and mutual influences between contexts, enabling it to obtain more informative context representations.
Similarly, user and item representations are obtained from their respective attributes (e.g., user gender, item category).
The core innovation of IEDR lies in the CIED module, which leverages the synergy between a context-invariant \textit{contrastive learning component} and a mutual information minimization-based \textit{disentangling component} to effectively differentiate intrinsic and extrinsic factors into disentangled representations. The contrastive learning component captures user preference that is stable across contexts by contrasting user representations under different contextual conditions. Concurrently, the disentangling component employs a bidirectional mutual information minimization scheme to separate the extrinsic factors that vary with different contexts from the intrinsic factors.
By jointly optimizing these two components, IEDR ensures that the learned intrinsic factors are not only stable across different contexts but also well-separated from the extrinsic factors. This innovative approach enables IEDR to effectively learn disentangled intrinsic and extrinsic factors, capturing the complex user behavior patterns for recommendation in various context scenarios.

In this paper, we make the following contributions:
\begin{itemize}[leftmargin=*]
\item We formally define intrinsic and extrinsic factors for recommender systems. Based on this definition, we propose IEDR, a novel framework that effectively learns intrinsic and extrinsic factors for more accurate recommendations. 
This is achieved by introducing two key components: a context-invariant contrastive learning component and a mutual information minimization-based disentangling component. These components work together to effectively capture the two types of factors from the interplay of various contexts.
%This is achieved by introducing a context-invariant contrastive learning component and a mutual information minimization-based disentangling component that works together to effectively capture the two types of factors from the interplay of various contexts. 
The implementation of IEDR is available at \url{https://github.com/ethanmock/IEDR}.

\item We theoretically analyze the proposed methods from an information theory perspective, providing insights into the effectiveness of our approach. We also identify key challenges and propose principled solutions to avoid degenerating results and ensure robust disentanglement, thereby improving recommendation accuracy and stability.

\item Extensive experiments on real-world datasets demonstrate that (1) IEDR significantly outperforms state-of-the-art methods by up to 4\% in NDCG, and (2) the proposed CIED module effectively learns disentangled intrinsic and extrinsic factors, leading to improved recommendation accuracy.
\end{itemize}

\section{Related Work}
\label{sec:relativeWork_factorDisentanglement}

This section summarizes the current research progress related to our work on factor disentanglement, feature interactions in recommender systems, and contrastive learning.

\subsection{Factor disentanglement}
Intrinsic and extrinsic factors are considered as two basic factors for individual decision-making in psychological research \cite{ryan2000intrinsic,benabou2003intrinsic,vallerand1997toward}. 
Recent recommender systems have borrowed the idea of capturing these two factors to achieve more accurate recommendations. 
For example, in the sequential recommendation, 
\citet{hidasi2016session} leverage the recurrent neural networks to capture users' long- and short-term (LS-term) interests from their interacted item sequences.
\citet{yu2019adaptive} propose a time-aware controller to capture the differences between LS-term interests for more accurate interest learning.
\citet{zheng2022disentangling} further emphasize the disentanglement between the LS-term interests at different time scales to differentiate the LS-term interests.
\citet{ning2023multi} demonstrate the effectiveness of embedding disentanglement by separating inter-domain and intra-domain knowledge.
%These models all focus on sequential recommendation, which can only capture the interests (factors) from time context.
\citet{wang2023causal} propose a Causal Disentangled Recommendation framework to handle user preference shifts by modeling the interaction generation procedure using a causal graph.
In point-of-interest recommendation, studies are leveraging spatial context to capture the intrinsic and extrinsic factors \cite{li2017learning,wu2020personalized}.
%In point-of-interest recommendation, IEMF \cite{li2017learning}, PLSPL \cite{wu2020personalized}.
However, all of the above studies focus on specific contexts. As a result, their factor learning approaches are hard to apply to other recommendation domains, which may result in a suboptimal solution if other contexts jointly influence these factors.
Some studies learn users' various factors without knowing the meaning of each factor (i.e., implicit factor). They first define the number of factors (e.g., 4) to be learned, and then disentangle the representations of each pair of factors \cite{ma2019learning,wang2020disentangled}. 
Compared to previous studies that focus on specific contexts or learn implicit factors, our IEDR model provides a generic framework to explicitly learn intrinsic and extrinsic factors from various contexts, enabling effective modeling of the complex interplay between stable user preference and various contextual influences in real-world recommendation scenarios.

\subsection{Feature interaction modeling} 
Many recommender systems leverage feature interactions to improve recommendation accuracy. 
One of the most common techniques is the factorization machine (FM) \cite{rendle2010factorization}, which models feature interactions through dot product and achieves great success. Recent studies extend FM with deep neural networks for more powerful feature interaction modeling \cite{xiao2017attentional,he2017neural,yu2019input,su2021neural}. The Wide \& Deep model (WDL) \cite{cheng2016wide} proposes a framework that combines shallow and deep modeling of features for recommendation. \cite{guo2017deepfm} combines FM and WDL by replacing the shallow part of WDL with an FM model. \cite{su2021detecting} leverages the relation reasoning power of graph neural networks for feature interaction modeling.
We are the first work to represent various contexts as a feature graph, and leverage graph neural networks to capture the interplay of the contexts in a feature interaction modeling paradigm for unified context learning.

\subsection{Contrastive learning} 
Contrastive learning has achieved great success in computer vision \cite{chen2020simple},
%\cite{chen2020simple,khosla2020supervised,chen2021exploring,fan2023contrastive,zhang2024vision} and 
neural language processing \cite{oord2018representation},
%\cite{oord2018representation,gao2021simcse,gunel2020supervised,xu2023making,sun2023contrastive}.
graph learning \cite{zhang2022benchmark,chen2023csgcl} and music learning \cite{yao2022contrastive}.
Recently, contrastive learning has attracted attention in recommender systems. 
\citet{yao2021self} conduct contrastive learning on users and items respectively on a two-tower framework to learn robust user and item representations.
In addition, \citet{wu2021self} propose a contrastive learning framework on a user-item bipartite graph to capture robust high-degree relationships between users and items.
\citet{ye2023towards} leverage contrastive learning on perturbed embeddings to improve the robustness of neural graph collaborative filtering.
\citet{wang2023sequential} propose a general framework called ContraRec that unifies two kinds of contrastive learning tasks, context-target contrast and context-context contrast, for sequential recommendation.
Some studies enhance recommendation through contrastive learning by mitigating popularity bias and promoting long-tail items with noise-based embedding augmentations \cite{yu2023xsimgcl,yu2022graph}. \citet{zhang2024empowering} propose AdvInfoNCE to handle false negatives and improve generalization. \citet{cai2023lightgcl} introduce LightGCL, using singular value decomposition to refine semantic structures and improve robustness. NCL incorporates structural and semantic neighbors as positive pairs for better user-item relationship learning \cite{lin2022improving}. 
%\citet{guo2024consistency} developed a contrastive tripartite graph learning method with consistency and discrepancy metrics to address cold-start scenarios.
The CETN model \cite{li2024cetn} addresses the challenge of capturing diverse and homogeneous feature interactions across semantic spaces by employing contrastive learning and self-supervised signals.
These works use contrastive learning to enhance recommendation by addressing bias, improving robustness, and promoting long-tail items.
Unlike previous works, we propose a context-invariant contrastive learning approach to capture stable intrinsic factors across various contexts, which is integrated with a mutual information minimization scheme to disentangle context-specific extrinsic factors.
%We propose the method that learns intrinsic factor representations that are invariant to context through a contrastive learning approach.

\section{Preliminary}

In this section, we introduce two key techniques that lay the foundation for our proposed method: the Statistical Interaction Graph Network (SIGN) \cite{su2021detecting} for effective feature interaction modeling, and the Variational Contrastive Log-ratio Upper Bound (vCLUB) \cite{cheng2020club} for mutual information estimation and minimization.

\subsection{Statistical Interaction Graph Network (SIGN)}
\label{sec:sign}

The statistical interaction graph network (SIGN) \cite{su2021detecting} explicitly models feature interactions through a graph neural network.
Given a set of features (e.g., user/item attributes) of each data sample, $\mathcal{Z}=\{z_1, z_2,...,z_n\}$, SIGN regards $\mathcal{Z}$ as a feature graph $\mathcal{G}(\mathcal{Z}, \mathcal{E})$, where $\mathcal{Z}$ is the node set that each feature $z_i$ is a node, and $\mathcal{E}$ is the edge set containing all the combinations of pairwise feature interactions, with each feature interaction $\langle z_i,z_j\rangle$ being an edge linking to corresponding nodes.
Accordingly, user representation learning becomes a graph learning problem.

In SIGN, first, each feature $z_i$ is mapped into a feature embedding $\bm{z}_i\in\mR^{d}$ of $d$ dimensions as the node embedding.
The embeddings are first randomly initialized and are updated through training.
Then, SIGN learns the graph representation (e.g., a vector) using a function $f$:
\begin{equation*}
\label{fun:sign}
    f(\mathcal{G}) = \phi(\{\psi(\{e_{ij}h(\bm{z}_{i}, \bm{z}_{j})\}_{j\in \mathcal{Z}})\})_{i \in \mathcal{Z}},
\end{equation*}
where $\phi$ and $\psi$ are aggregation functions (e.g., element-wise mean), $h(\cdot):\mR^{2\times d}\to\mR^{d}$ is an MLP that models each feature interaction, $e_{ij}\in\{0,1\}$ is the edge indicator (since we use all pairwise feature interactions, $e_{ij}=1$ for all edges).
$f$ outputs the modeled graph representation $\urep\in\mR^{d}$ of $d$ dimensions.

\subsection{Variational Contrastive Log-ratio Upper Bound (vCLUB) of Mutual Information}
\label{sec:club}

Given a set of sample pairs $\{(A_i,B_i)\}_{i=1}^N$ drawn from an unknown distribution $p(A,B)$ of random variables $A$ and $B$. The vCLUB method \cite{cheng2020club} derives the upper bound of their mutual information $\I(A,B)$ as:
\begin{equation}\label{eq:vCLUB_upper}
\begin{split}
  \I_{\text{vCLUB}}(A;B):= \mathbb{E}_{p(A,B)}[\log q_{\theta}(A|B)]
   - \mathbb{E}_{p(A)p(B)}[\log q_{\theta}(A|B)],  \\
\end{split}
\end{equation}
where $p(A,B)$ is the joint distribution, $p(A)p(B)$ is the marginal distribution, $q_{\theta}(A|B)$ is a variational distribution of parameter $\bm{\theta}$ (e.g., an MLP) to predict $A$ given $B$. 

In an application of mutual information minimization, we aim to reduce the correlation between $A_i$ and $B_i$ by selecting an optimal parameter $\bm{\sigma}$ of the joint variational distribution $p_{\sigma}(A,B)$.
vCLUB performs mutual information estimation and minimization in two steps iteratively. 
In the first step, to ensure Equation (\ref{eq:vCLUB_upper}) holds as the upper bound, 
$\bm{\theta}$ is trained to make the log-likelihood function $\mathcal{L}(A,B):=\frac{1}{N}\sum^{N}_{i=1}\log q_{\theta}(A_i|B_i)$ maximized (Theorem 3.2 of \cite{cheng2020club}).
In the second step, $\bm{\theta}$ is frozen, and other parameters ($\bm{\sigma}$) are trained to minimize $\I_{\text{vCLUB}}(A;B)$ so that the mutual information is minimized.

\iffalse
Suppose we aim to learn an intrinsic representation $\Uin$ and an extrinsic representation $\Uex$, where their mutual information is minimized. 
In the vCLUB method \cite{cheng2020club}, a variational distribution $q^{u}_{1}(\Uex|\Uin;\bm{\theta}_1^{u})$ of parameter $\bm{\theta}_1^{u}$ (e.g., an MLP) is proposed to predict the extrinsic factor given the intrinsic factor.
Then, the vCLUB-based mutual information upper bound can be derived as:
\begin{equation}\label{eq:vCLUB_upper}
\small
\begin{split}
  \I_{\text{vCLUB}}(\Uin;\Uex):=& \mathbb{E}_{p(\Uin,\Uex)}[\log q_{1}^{u}(\Uex|\Uin)] \\
   - &\mathbb{E}_{p(\Uin)p(\Uex)}[\log q_{1}^{u}(\Uex|\Uin)]  \\
\end{split}
\end{equation}
where $p(\Uin,\Uex)$ is the joint distribution and $p(\Uin)p(\Uex)$ are is the marginal distribution.

vCLUB performs mutual information estimation and minimization in two steps iteratively. 
In the first step, to ensure Equation (\ref{eq:vCLUB_upper}) holds as the upper bound, $\bm{\theta}_1^{u}$ is trained to make the log-likelihood function $\mathcal{L}_{appr}(u,c):=\frac{1}{N}\sum^{N}_{i=1}\log q_{1}^{u}((\Uex)_i|(\Uin)_i)$ maximized (Theorem 3.2 of \cite{cheng2020club}). 
In the second step, $\bm{\theta}_1^{u}$ is frozen, and other parameters (e.g., the parameters to generate $\Uin$ and $\Uex$) are trained to minimize $\I_{\text{vCLUB}}(\Uin;\Uex)$ so that the mutual information is minimized.
\fi

\section{Problem Statement and Definitions}
Let $\mathcal{U}$, $\mathcal{V}$, and $\mathcal{C}$ denote the user set, item set, and context set, respectively.
Each user $u\in \mathcal{U}$ consists a set of user features $u=\{z^{u}_1, z^{u}_2,...,z^{u}_p\}$ %that describe the user 
(e.g., user ID, gender). 
%Note that the feature number $p$ may vary for different users
Similarly, each item $v\in\mathcal{V}$ is represented by a set of item features $v=\{z^{v}_1, z^{v}_2,...,z^{v}_q\}$ (e.g., branch, color). % that describe the item.
A context $c\in\mathcal{C}$ is a set of context features $c=\{z^{c}_1, z^{c}_2,...,z^{c}_m\}$, denoting the context state when a user selects an item (e.g., weather, daytime).
Let $\mathcal{D}$ be a dataset containing $N$ instances (i.e., data samples) of $(u, v, c)$, with a corresponding label $y\in\{1,0\}$ indicating whether or not the user $u$ selects the item $v$ under the context $c$. 
The recommendation task can be formulated as predicting the selection probability $y^{\prime}=p(u, v, c)$.
In our proposed IEDR model, the intrinsic factor $\Gin$ and the extrinsic factor $\Gex$ are explicitly inferred for both users and items, and jointly leveraged to perform the prediction.

Next, we formally define intrinsic and extrinsic factors. We believe these two factors exist from both users' and items' perspectives. 
This is reasonable since a user selecting an item not only relates to the factors (motivations) of users, e.g., prefer \textit{healthy food} (intrinsic factor) on \textit{weekdays} (extrinsic factor), but also relates to the factors (attractiveness) of items, e.g., the Caesar salad is \textit{healthy} (intrinsic factor) and is chosen more often when \textit{the weather is hot} (extrinsic factor).
In the following, we define intrinsic and extrinsic factors from the users’ perspective only, as they are similar from the items' perspective.

\begin{definition}
\label{def:in_ex_factor}
(\textbf{Intrinsic Factor and Extrinsic Factor}) 
Consider a user $u$ and a set of contexts $\mathcal{C}$; 
an \textbf{intrinsic factor} of the user is a factor that is 
%a factor is called user \textbf{intrinsic factor} if it is 
invariant to the contexts in $\mathcal{C}$, i.e., $f_{in}(u,c)=f_{in}(u,c^{\prime})$, where $f_{in}$ is a function learning intrinsic factor representations, and $c$ and $c^{\prime}$ are two arbitrary contexts in $\mathcal{C}$.
On the other hand, 
an \textbf{extrinsic factor} of the user is a factor that 
is different from its corresponding intrinsic factor, i.e., $\I(f_{in}(u,c), f_{ex}(u,c))=0$, where $\I$ computes the mutual information and $f_{ex}$ learns extrinsic factor representations. Also, the extrinsic factor changes w.r.t. the context, i.e., there exist contexts $c$ and $c^{\prime}$ in $\mathcal{C}$ such that $f_{ex}(u,c) \neq f_{ex}(u,c^{\prime})$.
\end{definition}

In the definition, $f_{in}(u,c)=f_{in}(u,c^{\prime})$ shows the invariance of intrinsic factors.%, e.g., different weather conditions ($c, c^{\prime}$) do not influence the user's preference on user interface design (intrinsic factor). 
On the other hand, $f_{ex}(u,c) \neq f_{ex}(u,c^{\prime})$ shows that the extrinsic factors can be different if the contexts are different.

In previous research (both in psychology \cite{benabou2003intrinsic,vallerand1997toward} and in recommender systems \cite{hidasi2016session,yu2019adaptive}), intrinsic and extrinsic factors are considered all the factors influencing user behavior, and learning these two factors in a disentangled way has proven effective to analyze these behaviors \cite{zheng2022disentangling}.
%Meanwhile, Definition \ref{def:in_ex_factor} demonstrates that the information extracted by $f_{in}$ and $f_{ex}$ should be mutually exclusive.
Therefore, it leads to our factor learning objective based on Definition \ref{def:in_ex_factor}:
leveraging the context-invariant property to ensure that $f_{in}$ captures intrinsic factors, and disentangling the outputs of $f_{in}(u,c)$ and $f_{ex}(u,c)$ to ensure $f_{ex}$ captures extrinsic factors (detailed in Section \ref{sec:CIED}).

\section{Intrinsic-Extrinsic Disentangled Recommendation Model}

\begin{figure*}[t]
\centerline{\includegraphics[width=0.94\textwidth]{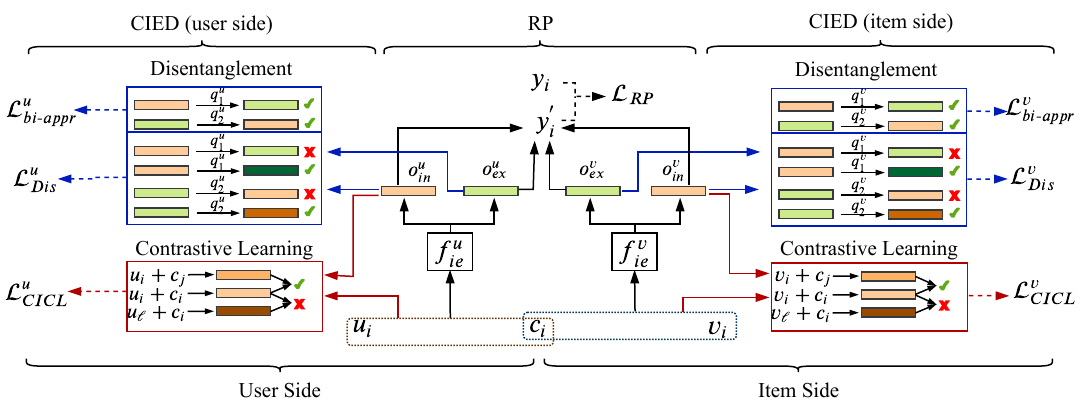}}
%\vskip -0.08in
\caption{ An Overview of IEDR.
It is a symmetric structure on the user side and the item side.
The middle part (the black arrows) represents the recommendation prediction (RP) module (Section \ref{sec:rpm}). It generates the intrinsic and extrinsic factor representations ($\Gin$ and $\Gex$) for producing the recommendation prediction $y^{\prime}$. 
The side parts are two contrastive intrinsic-extrinsic disentangling (CIED) modules. Each CIED includes a context-invariant contrastive learning component (the red arrows, Section \ref{sec:method_CICL}), and a disentangling component (the blue arrows, Section \ref{sec:method_biDis}) to ensure the success of the factor learning.
The losses generated through these modules ($\mathcal{L}_{RP},\mathcal{L}_{CICL},\mathcal{L}_{bi\text{-}appr},\mathcal{L}_{Dis}$) will be optimized as a two-step multi-task training (Section \ref{sec:method_multitask}).}
\Description{An Overview of IEDR}
\label{fig:model_overview}
\end{figure*}

To effectively learn and disentangle intrinsic and extrinsic factors from various contexts, we propose our Intrinsic-Extrinsic Disentangled Recommendation (IEDR) Model.
The overview of our model is visualized in Figure \ref{fig:model_overview}. 
More specifically, our proposed IEDR model consists of the following two modules, which will be detailed in the next subsections:
\begin{itemize}[leftmargin=*]
    \item A recommendation prediction (RP) module that takes a user and an item as input, and combines them with a set of contexts, to generate intrinsic and extrinsic factor representations for both the user and the item.
    %that takes user, item and context information as input, to generate intrinsic and extrinsic factor representations for both users and items, respectively. 
    The predicted probability $y^{\prime}$ is then jointly learned from these representations.
    \item A contrastive intrinsic-extrinsic disentangling (CIED) module is applied to both the user and the item sides to support the intrinsic and extrinsic factor learning. 
    The module contains a context-invariant contrastive learning component and a disentangling component, to ensure the learned factors satisfy Definition \ref{def:in_ex_factor}.
    %The module contains two components: a context-invariant contrastive learning component, and a disentangling component. Together, the two components ensure the learned factors satisfy Definition \ref{def:in_ex_factor}. 
\end{itemize}

\noindent
For clarity and ease of understanding, Table~\ref{tab:notations} summarizes the key notations used throughout the IEDR model.

\begin{table}[t]
\centering
\caption{Summary of notations used in the IEDR model.}
\label{tab:notations}
{
\begin{tabular}{ll}
\toprule
\textbf{Notation} & \textbf{Description} \\
\midrule
$U, V, C$ & Sets of users, items, and contexts, respectively. \\
$\bm{z}^u_i, \bm{z}^v_i, \bm{z}^c_i$ & The $i^{th}$ feature representation of user $u$, item $v$, and context $c$. \\
$\bm{u}$, $\bm{v}$, $\bm{c}$ & User, item, and context representations. \\
$\bm{o}_\text{in}, \bm{o}_\text{ex}$ & Intrinsic and extrinsic factor representations. \\
%$f_u, f_v, f_c$ & Functions to learn user, item, and context representations. \\
%$f_\text{ie}$ & Function to generate intrinsic and extrinsic factors. \\
$\mathcal{L}_\text{RP}$ & Recommendation prediction loss. \\
$\mathcal{L}_\text{CICL}$ & Context-invariant contrastive learning loss. \\
$\mathcal{L}_\text{bi-appr}$ & Bidirectional approximation loss for disentanglement. \\
$\mathcal{L}_\text{Dis}$ & Disentanglement loss. \\
\bottomrule
\end{tabular}
}
\end{table}

%Next, we introduce each of these modules, followed by the theoretical analysis of our model. 
%In the following subsections, we will detail the two modules.

\subsection{Recommendation Prediction (RP) Module}
\label{sec:rpm}

The recommendation prediction (RP) module is a symmetric structure that generates user intrinsic and extrinsic factor representations ($\Uin, \Uex$) from the user side, and generates item intrinsic and extrinsic factor representations ($\Iin, \Iex$) from the item side. %, respectively.
On the user side,
we first generate a user representation and a context representation based on user features and context features, respectively.
Here, we use the SIGN model \cite{su2021detecting} to generate the representations (see Section \ref{sec:sign} for details). SIGN has been proven effective in user/item/context representation learning through modeling feature interactions via graph neural networks. 
%Appendix \ref{appx:sign} provides a detailed description of SIGN.
More formally, let $f_u(u):\mR^{p\times d}\to\mR^{d}$ be the function for SIGN-based feature modeling, where $p$ is the number of user features.
$f_u(u)$ first maps each user feature $z^{u}_i\in u$ into a $d$-dimensional feature embedding $\bm{z}^{u}_{i}$. 
Then, it models these feature embeddings to output the user representation $\urep$.
Similarly, SIGN learns context representation $\crep$ through $f_c$.
Next, a factor generation function $f_{ie}^{u}(\urep, \crep):\mR^{2\times d}\to\mR^{2\times d}$ (e.g., a neural network) takes the user representation and the context representation as input, and simultaneously generates a user intrinsic representation $\Uin$ and a user extrinsic representations $\Uex$.
Here, the output is a $2d$-dimensional vector, with the first $d$-dimensional terms as $\Uin$ and the rest as $\Uex$. %last $d$ dimensions as $\Uex$.
Note that without our CIED module (Section \ref{sec:CIED}), $\Uin$ and $\Uex$ are entangled. Currently, we name them $\Uin$ and $\Uex$ to make it consistent with the following description. When equipped with CIED module, $\Uin$ and $\Uex$ will be disentangled and represent intrinsic and extrinsic factors respectively.
On the item side, a similar module structure is adopted.
%Also, the same model structure is adopted on the item side. 
We use a different SIGN-based function for the item representation learning $\irep=f_v(v)$, while using the same context representation as that on the user side. A factor-generating function $f_{ie}^{v}(\irep, \crep)$ is applied to obtain the item intrinsic factor representation $\Iin$ and extrinsic factor representation $\Iex$.

Finally, we learn the prediction $y^{\prime}=f_{pred}(\Uin,\Uex, \Iin, \Iex)$. 
We linearly combine the learned factors and use the dot product as the prediction function: % to perform the user-item matching (prediction): 
$f_{pred}(\Uin, \Uex, \Iin, \Iex) = (\Uin+\Uex)^{\top}(\Iin+\Iex)$.
A cross-entropy loss function is adopted to minimize the prediction error: %the error between the prediction and the true label: 
$\mathcal{L}_{\text{RP}}(u, v, c):= - y\log(y^{\prime}) + (1-y)\log(1-y^{\prime})$.

\subsection{Contrastive Intrinsic-Extrinsic Disentangling (CIED) Module}
\label{sec:CIED}
The CIED module is designed to capture intrinsic and extrinsic factors from the representations generated by the RP module. The key idea is to integrate a context-invariant contrastive learning objective with a mutual information minimization scheme to simultaneously capture intrinsic factors that are stable across contexts and extrinsic factors that vary with different contextual conditions.

Specifically, CIED consists of two interrelated components: (1) a context-invariant contrastive learning component that encourages the model to learn intrinsic factors by contrasting user representations across different contexts, and (2) a bidirectional disentangling component that further separates the extrinsic factors from the learned intrinsic factors via a bidirectional mutual information minimization scheme. Next, we describe the two components in detail. 

\subsubsection{Context-Invariant Contrastive Learning Component}
\label{sec:method_CICL}

The context-invariant contrastive learning component is designed to learn intrinsic representations that are invariant across different contexts. The core idea is to maximize the agreement between the intrinsic representation pairs generated from the same user under different contexts (positive pairs), while minimizing the agreement between those generated from different users under the same context (negative pairs). This contrastive objective encourages the model to capture the shared information across contexts as the intrinsic representation.
More formally, we represent the intrinsic representations with the subscript $(\Uin)_{ij}$ if it is generated through user $u_i$ (from $i$-th data sample) and context $c_j$ (from $j$-th data sample), i.e., $(\Uin)_{ij}=f_{ie}^{u}(\urep_i, \crep_j)$.
%For easy understanding, we use subscript to describe the user-context combination that generates the intrinsic representation, e.g., $(\Uin)_{ij}=f_{ie}^{u}(u_i, c_j)$.
Inspired by InfoNCE \cite{oord2018representation}, for the $i$-th data sample $(u_i, v_i, c_i)\in\mathcal{D}$, %we calculate an InfoNCE-based \cite{oord2018representation} objective function $\mathcal{L}_{\text{CICL}}^{u}$:
we calculate the objective function as follows:  
\begin{equation}\label{eq:CICL_loss_CL}
\mathcal{L}_{\text{CICL}}^{u}(u_i, c_i) := -\log \frac{\exp\left(\text{sim}((\Uin)_{ii}, (\Uin)_{ij})/\tau\right)}{\sum_{u_\ell\in\mathcal{U}}\exp\left(\text{sim}((\Uin)_{ii}, (\Uin)_{\ell i})/\tau\right)},
\end{equation}
where $(\Uin)_{ij}$ is generated from a user $u_i$ and an arbitrary context $c_j$, $\text{sim}(\cdot)$ is the cosine similarity, and $\tau$ is a temperature value.

The objective function is intuitive: one user should have the same intrinsic factor in different contexts, while different users can have their own personalized interests (different intrinsic factors).

\subsubsection{Disentangling Component}
\label{sec:method_biDis}

To capture both the intrinsic and extrinsic factors, we need to disentangle extrinsic factors from intrinsic factors.
The vCLUB method \cite{cheng2020club} can perform disentanglement through mutual information minimization. However, typical vCLUB is an asymmetric method, which may be less robust and lead to unsatisfactory disentanglement (detailed in Section \ref{appx:vclub_problem}).
Therefore, we propose a bidirectional vCLUB approach that simultaneously minimizes the mutual information between intrinsic and extrinsic factors in both directions, leading to more robust and effective disentanglement.

In the bidirectional vCLUB, two variational distributions (e.g., approximated via neural networks)
$q_{1}^{u}(\Uex|\Uin;\bm{\theta}_1^u)$ and $q_{2}^{u}(\Uin|\Uex; \bm{\theta}_2^u)$ are proposed with parameters $\bm{\theta}_1^u$ and $\bm{\theta}_2^u$, to predict the two types of factors, respectively.
Then a bidirectional vCLUB-based mutual information upper bound can be obtained as:\footnote{$\I_{\text{bi-vCLUB}}(\Uin;\Uex)$ is the average of two vCLUB-based upper bounds of different directions. Therefore, it is obvious that $\I_{\text{bi-vCLUB}}(\Uin;\Uex)$ is still an upper bound of $\I(\Uin;\Uex)$.}
\begin{equation}\label{eq:bi-vCLUB_upper}
\begin{split}
    \I_{\text{bi-vCLUB}}(\Uin;\Uex):= &
    \frac{1}{2}\Bigl(\mathbb{E}_{p(\Uin,\Uex)}[\log q_{1}^{u}(\Uex|\Uin)] - \mathbb{E}_{p(\Uin)p(\Uex)}[\log q_{1}^{u}(\Uex|\Uin)]  \\
    +& \mathbb{E}_{p(\Uin,\Uex)}[\log q_{2}^{u}(\Uin|\Uex)] - \mathbb{E}_{p(\Uex)p(\Uin)}[\log q_{2}^{u}(\Uin|\Uex)]\Bigr). \\
\end{split}
\end{equation}

By minimizing the upper bound $\I_{\text{bi-vCLUB}}(\Uin;\Uex)$ as above, we minimize the mutual information between $\Uin$ and $\Uex$.
Experimental results in Section \ref{sec:vCLUBvsbiDis} show that vCLUB is more robust and achieves better factor learning.

The optimization of the disentangling component is conducted in two iteratively steps.
In the first step, we estimate the upper bound by training $\bm{\theta}_1^{u}$ and $\bm{\theta}_2^{u}$ to minimize the loss function $\mathcal{L}_{bi\text{-}appr}^{u}(u_i,c_i):=-\frac{1}{2}\Bigl(\log q_{1}^{u}\bigl((\Uex)_{ii}|(\Uin)_{ii}\bigr)+\log q_{2}^{u}\bigl((\Uin)_{ii}|(\Uex)_{ii}\bigr)\Bigr)$. 
Following \cite{cheng2020club}, we use the mean squared error to optimize $q_{1}^{u}$ and $q_{2}^{u}$.
%We use mean squared error to optimize $q_{1}^{u}$ and $q_{2}^{u}$.
In the second step, we freeze $\bm{\theta}_1^{u}$ and $\bm{\theta}_2^{u}$, and minimize the mutual information of $\Uin$ and $\Uex$ by training other parameters to minimize the upper bound $\mathcal{L}_{Dis}^{u}(u_i, c_i)=\I_{\text{bi-vCLUB}}\bigl((\Uin)_{ii};(\Uex)_{ii}\bigr)$.
%\subsubsection{Integration of Contrastive Learning and Disentanglement}

The context-invariant contrastive learning and disentanglement components in CIED are designed to work synergistically to learn meaningful intrinsic and extrinsic factors in the recommendation setting of various contexts. The contrastive learning component first learns context-invariant intrinsic factors by contrasting user representations across different contexts. These learned intrinsic factors then serve as a starting point for the disentanglement component to further separate the extrinsic factors via bidirectional mutual information minimization.

The seamless integration of these two components is crucial for the effectiveness of IEDR. By first learning context-invariant factors and then disentangling them from the extrinsic factors, CIED can effectively capture the complex user behavior patterns influenced by various contextual conditions. Unlike existing methods, IEDR ensures context-agnostic learning of intrinsic and extrinsic factors in recommendations in scenarios of various contexts, and uniquely considers the interplay between these factors across various contexts, enhancing the model's effectiveness in complex, dynamic recommendation scenarios.

\subsection{Implementation Details}
\label{sec:implementation}

\subsubsection{Iterative Optimization Procedure}
\label{sec:method_implementation}

The CIED module is implemented as an iterative optimization procedure that alternates between the context-invariant contrastive learning and the disentanglement components.

In each iteration, the contrastive learning component first updates the model parameters to learn context-invariant intrinsic factors. Specifically, for each user $u_i$ and context $c_i$ in the current batch, we generate a positive pair $(\Uin)_{ij}$ by either (1) randomly sampling a context $c_j$ from the same batch, or (2) applying a high dropout rate to the original context representation $c_i$. We also generate $L$ negative pairs $(\Uin)_{\ell i}$ by randomly sampling $L$ users from the same batch. The contrastive loss $\mathcal{L}_{\text{CICL}}^{u}(u_i, c_i)$ (Equation \ref{eq:CICL_loss_CL}) is then computed and minimized to update the model parameters.

The learned intrinsic factors $(\Uin)_{ii}$ are then fed into the disentangling component, which estimates and minimizes the mutual information between the intrinsic and extrinsic factors. We introduce two variational distributions $q{1}^{u}(\Uex|\Uin;\bm{\theta}1^u)$ and $q{2}^{u}(\Uin|\Uex; \bm{\theta}_2^u)$, parameterized by $\bm{\theta}_1^u$ and $\bm{\theta}2^u$, to estimate the bidirectional mutual information upper bound $\I{\text{bi-vCLUB}}(\Uin;\Uex)$ (Equation \ref{eq:bi-vCLUB_upper}). The disentangling component is optimized in a two-step procedure: (1) estimating the mutual information upper bound by optimizing $\bm{\theta}1^{u}$ and $\bm{\theta}2^{u}$ to minimize the loss $\mathcal{L}{bi\text{-}appr}^{u}(u_i,c_i)$, and (2) minimizing the mutual information by optimizing the other parameters to minimize the upper bound $\mathcal{L}{Dis}^{u}(u_i, c_i)$.

The updated extrinsic factors $(\Uex)_{ii}$ are then used to refine the intrinsic factors in the next iteration of contrastive learning. This iterative process continues until convergence or a maximum number of iterations is reached.

%This iterative optimization procedure ensures that the learned intrinsic and extrinsic factors are maximally informative of the user's stable preferences and contextual influences while being minimally dependent on each other. The tight integration of contrastive learning and mutual information minimization enables CIED to effectively capture the complex user behavior patterns in the multi-context recommendation setting. 

\subsubsection{Multi-task Training}
\label{sec:method_multitask}

We perform a two-step multi-task training to minimize the empirical risk of multiple components in IEDR. %minimization function of IEDR.
The two steps run alternatively until convergence. % the model is convergent.
Appendix \ref{appx:algorithm} provides the pseudo-code of the training procedure.
In the first step, we freeze all the parameters except for $\bm{\theta}_1^{u}, \bm{\theta}_2^{u}, \bm{\theta}_1^{v}$, and $\bm{\theta}_2^{v}$, where $\bm{\theta}_1^{v}, \bm{\theta}_2^{v}$ are the parameters of $q_{1}^{v}(\Iex|\Iin; \bm{\theta}_1^{v})$ and $q_{2}^{v}(\Iin|\Iex; \bm{\theta}_2^{v})$ in the disentangling component on the item side.
We then minimize $\mathcal{R}(\bm{\theta}_1^{u}, \bm{\theta}_2^{u}, \bm{\theta}_1^{v}, \bm{\theta}_2^{v}) =\frac{1}{N}\sum_{i=1}^{N} \bigl(\mathcal{L}_{bi\text{-}appr}^{u}(u_i,c_i) + \mathcal{L}_{bi\text{-}appr}^{v}(v_i,c_i)\bigr)$. In the second step, we freeze $\bm{\theta}_1^{u}, \bm{\theta}_2^{u}, \bm{\theta}_1^{v}$, and $\bm{\theta}_2^{v}$, and minimize the following function:

\begin{equation*}\label{eq:total_loss}
\begin{split}
 \arg\min \mathcal{R}(\bm{\omega}) \!=\! \frac{1}{N}\sum_{i=1}^{N} \Bigl(\mathcal{L}_{\text{RP}}(u_i, v_i, c_i)
+ \lambda_1 \bigl(\mathcal{L}_{\text{CICL}}^{u}(c_i, u_i)  + \mathcal{L}_{\text{CICL}}^{v}(c_i, v_i)\bigr) + \lambda_2 \bigl(\mathcal{L}_{Dis}^{u}(u_i, c_i) \!+\! \mathcal{L}_{Dis}^{v}(v_i, c_i)\bigr)\Bigr), \\
\end{split}
\end{equation*}
where $\mathcal{L}_{bi\text{-}appr}^{v}$, $\mathcal{L}_{\text{CICL}}^{v}$, and $\mathcal{L}_{Dis}^{v}$ are the losses on the item side, $\lambda_1$ and $\lambda_2$ are the weight factors, and $\bm{\omega}$ are all the trainable parameters except for $\bm{\theta}_1^{u}, \bm{\theta}_2^{u}, \bm{\theta}_1^{v}$, and $\bm{\theta}_2^{v}$.

The multi-task training procedure ensures that the model learns to accurately predict recommendations while simultaneously learning disentangled intrinsic and extrinsic factors. The contrastive learning and disentanglement losses are integrated into the overall training objective, allowing the model to capture the complex user behavior patterns influenced by various contextual conditions.

%By combining the iterative optimization procedure of the CIED module with the multi-task training strategy, IEDR can effectively learn and disentangle intrinsic and extrinsic factors from the representations generated by the RP module, leading to improved recommendation performance in the multi-context setting.

\section{Discussion}
In this section, we provide theoretical and practical discussions of IEDR from multiple perspectives, including the information theory foundation, time complexity analysis, trivial solution prevention, and potential problems of the vCLUB method used in the disentanglement component.

\subsection{Theoretical Analysis: Context-invariant Contrastive Learning in Information Theory}

In this section, we reason the context-invariant contrastive learning from the perspective of information theory.
%We show that 
As formally defined in Theorem \ref{thm:theoryCICL}, optimizing Equation (\ref{eq:CICL_loss_CL}) is equivalent to maximizing the mutual information between the intrinsic representations and user representations, and simultaneously minimizing the mutual information between the intrinsic representations and the context representations. The theorem on the item side can be derived in the same fashion.
%we analyze the equivalence of optimizing CICL to mutual information maximization and minimization. 
The proof of this equivalence can be found in Appendix \ref{appx:prooftheoryCICL}.
%For simplicity, we take the module on the user side as a representative demonstration, and the theory on the item side can be inferred in the same way.
%Here, we derive the theorem for the user side, and the item side's can be derived in the same way.%fashion.

\begin{theorem}[Equivalence of contrastive loss $\mathcal{L}_{\textit{CICL}}^{u}$]
\label{thm:theoryCICL}
Optimizing the contrastive loss is equivalent to solving:
\begin{equation}
\label{eq:CICLeqInfomaxmin}
\arg\!\min  \!\sum_{i=1}^{N}\mathcal{L}_{\text{CICL}}^{u}(u_i,\! c_i) \!=\!  \arg \max \Bigl(\I(\Uin, \!\urep) \!- \! \I(\Uin, \!\crep)\Bigr).
\end{equation}
\end{theorem}

Theorem \ref{thm:theoryCICL} provides the perspective from information theory to understand the context-invariant contrastive learning procedure: the information of users that is not influenced by contexts (i.e., intrinsic factors) is kept in $\Uin$.
%More intuitively, the contrastive self-supervised module is equivalent to maximizing the mutual information (InfoMax \citep{linsker1988infomax}) between the intrinsic user interest and user features, and minimizing the mutual information (InfoMin \citep{tian2020makes}) between the intrinsic user interest and the context information, simultaneously.

\subsection{Time Complexity Analysis}
\label{sec:timeComplexityAnalysis}

The time complexity of IEDR is comparable to feature interaction-based recommender systems (e.g., AutoInt \cite{song2019autoint}, SIGN \cite{su2021detecting}). The overhead of the alternative optimizing procedure for the disentanglement component is marginal in the whole optimizing procedure.

Specifically, the most time-consuming computations are the feature interaction learning to get user, item, and context representations, which need to conduct interaction modeling on every pair of feature interactions. This procedure has also been done on other feature interaction-based models. Therefore, the time complexity of the proposed module is comparable with those methods.

Our model takes additional computations on the contrastive learning component (CICL) and the disentangling component: (1) For the CICL component, we do not need to perform the feature interaction modeling again, but reuse the generated user, item and context representations, which saves the majority of the overhead. We only need to perform $f_{ie}$ $L+1$ times, where $L$ is the number of negative samples and $f_{ie}$ is a one-hidden layer MLP. (2) For the disentangling component, we reuse the generated user/item/context representations as well. The first step in the two-step learning takes very little overhead. This is because this step only tries to optimize the parameters of the functions $q_1$ and $q_2$ (Equation (\ref{eq:bi-vCLUB_upper})), which are two MLPs with one hidden layer. For each data sample, we only run $q_1$ and $q_2$ once using $\Gin$ and $\Gex$.

In summary, since all of the computations above do not need to perform feature interaction modeling (the most time-consuming procedure in all feature interaction-based models), the small imposed overhead is acceptable considering the effectiveness of our model in capturing accurate intrinsic and extrinsic factors. More empirical analysis can be found in Section \ref{sec:Emperical_timeComplexity}.

\subsection{Preventing the Trivial Solution of CIED}
\label{appx:trivial_solution}

The two components in the CIED module, the contrastive learning component and the disentangling component, jointly ensure the success of the intrinsic and extrinsic factor representation learning. 
However, CIED may fall into a trivial solution: $f_{ie}^{u}(\urep, \crep)$ maps $u$ to $\Uin$ without considering $c$, and maps $c$ to $\Uex$ without considering $u$.
Although this trivial solution minimizes $\mathcal{L}_{\text{CICL}}(u, c)$ and $\mathcal{L}_{Dis}(u, c)$,
%As a result 
$\Uin$ (resp. $\Uex$) is not the intrinsic (resp. extrinsic) factor, but just a mapping of the user information (resp. context information).
We prove that this trivial solution can be avoided by setting $f_{ie}^{u}(\urep, \crep)$ as a \textit{non-linear} function, leading $\urep$ and $\crep$ to statistically interact.

\subsubsection{Statistical Interaction}

We first introduce the statistical interaction (or non-additive interaction), which ensures a joint influence of several variables on an output variable is not additive \cite{tsang2018neural}.
Based on \cite{sorokina2008detecting}, $F(\bm{X})$ shows statistical interaction between variables $x_i$ and $x_j$ if $\forall f_{\char`\\i}, f_{\char`\\j}$, $F(\bm{X})$ \textbf{cannot} be expressed as:
%the sum of the functions $f_{\char`\\i}$ and $f_{\char`\\j}$:

\begin{equation}
\label{fun:spi}
\begin{split}
 F(\bm{X}) \neq &f_{\char`\\i}(x_{1},\ldots,x_{i-1},x_{i+1},\ldots,x_{n}) + f_{\char`\\j}(x_{1},\ldots,x_{j-1},x_{j+1},\ldots,x_{n}).  \\
\end{split}
\end{equation}

More generally, if using $\bm{v}_{i}\in\mathbb{R}^{d}$ to describe the $i$-th variable with a $d$-dimension vector \cite{rendle2010factorization,su2021detecting}, e.g., variable embedding, each variable can be described in a vector form $\bm{u}_i=x_i\bm{v}_i$. 
Then, we define the pairwise statistical interaction in vector form by changing the Equation (\ref{fun:spi}) into:  
\begin{equation*}
\label{eq:spi_emb}
\begin{split}
  F(\bm{X}) \neq &f_{\char`\\i}(\bm{u}_1,\ldots,\bm{u}_{i-1},\bm{u}_{i+1},\ldots,\bm{u}_{n}) + f_{\char`\\j}(\bm{u}_{1},\ldots,\bm{u}_{j-1},\bm{u}_{j+1},\ldots,\bm{u}_{n}).
\end{split}
\end{equation*}

\subsubsection{Preventing the Trivial Solution}
%As stated in Section \ref{sec:theory_trivialsolution}, the trivial solution of CIED appears if $f_{ie}^{u}(\urep, \crep)$ trivially maps $\urep$ to $\Uin$ without considering $\crep$, and maps $\crep$ to $\Uex$ without considering $\urep$. 
Based on the definition of statistical interaction, we can express the trivial solution as that $f_{ie}^{u}(\urep, \crep)$ learns no statistical interaction between $\urep$ and $\crep$:
\begin{equation}
\label{eq:trivial_sol}
    f_{ie}^{u}(\urep, \crep) = \lambda_1 f_{1}(\urep) + \lambda_2 f_{2}(\crep),
\end{equation}
where $f_{1}$ outputs $\Uin$, $f_{2}$ outputs $\Uex$, and $\lambda$ are weight scalars. 

To prevent the trivial solution, we need to ensure that function $f_{ie}^{u}(\urep, \crep)$ cannot be modeled in the form of Equation (\ref{eq:trivial_sol}). Therefore, if $\urep$ and $\crep$ are modeled as a statistical interaction in $f_{ie}^{u}(\urep, \crep)$, the trivial solution can be prevented.
Since $f_{ie}^{u}(\urep, \crep)$ only takes $\urep$ and $\crep$ as inputs, we just need $f_{ie}^{u}$ to be a non-additive model. That is, $f_{ie}^{u}(\urep, \crep)$ should contain a third term $f_{3}(\urep, \crep)$:
\begin{equation*}
\label{eq:non_trivial_sol}
    f_{ie}^{u}(\urep, \crep) = \lambda_1 f_{1}(\urep) + \lambda_2 f_{2}(\crep) + \lambda_3 f_{3}(\urep, \crep),
\end{equation*}
where $f_{3}$ is a non-additive model and $\lambda_3\neq 0$.

Therefore, in the optimized situation, $\Uin=\lambda_1 f_{1}(\urep)$ learns part of the information from users that do not interact with context information. 
$\Uex=\lambda_2 f_{2}(\crep) + \lambda_3 f_{3}(\urep, \crep)$ learns the context information ($f_{2}(\crep)$) and the information that changes given different contexts ($f_{3}(\urep, \crep)$).

In Section \ref{appx:exp_trivial}, we empirically analyze how the trivial solution will influence the prediction performance.

\subsection{Potential Problems of the Asymmetric vCLUB Method}
\label{appx:vclub_problem}

The vCLUB-based mutual information minimization method proposed in \cite{cheng2020club} is an asymmetric method. 
%Appendix \ref{appx:club} gives an introduction about how vCLUB method performs mutual information minimization.
In this section, we explain the possible reason that \textit{vCLUB} is less robust and performs worse than our proposed bidirectional vCLUB method (\textit{BiDis}). 

\begin{figure}[t]
\centerline{\includegraphics[width=0.75\textwidth]{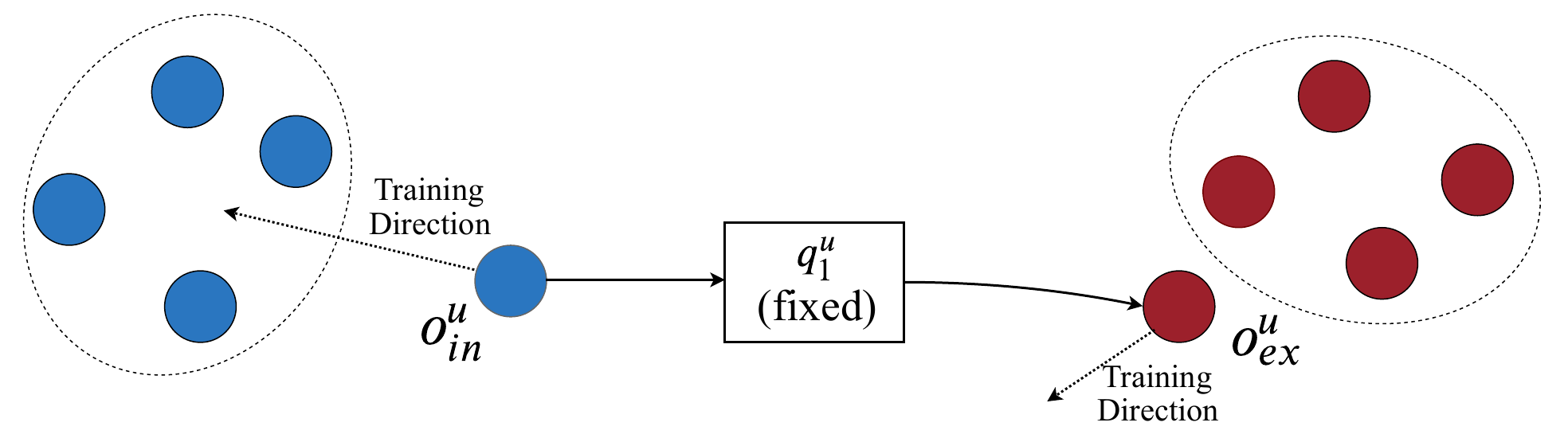}}
\caption{ An illustrative example demonstrating the potential problem of asymmetric learning in vCLUB. The blue circles are intrinsic representations, and the red circles are extrinsic representations. The dotted arrows are the directions that \textit{vCLUB} will push $\Uin$ and $\Uex$ to move toward their space.}
\Description{An illustrative example demonstrating the potential problem of asymmetric learning in vCLUB.}
\label{fig:sinDis_problem}
\end{figure}

Directly applying vCLUB leads to the parameter $\bm{\theta}_1^u$ of a variational distribution $q^u_{1}(\Uex|\Uin;\bm{\theta}_1^u)$ being trained to approach the vCLUB-based upper bound in Equation (\ref{eq:vCLUB_upper}) (Step 1).
Then, $\bm{\theta}_1^u$ is frozen, and $\Uex, \Uin$ are trained to minimize $\I(\Uin;\Uex)$ via minimizing the upper bound $\I_{\text{vCLUB}}(\Uin;\Uex)$ (Step 2). 
However, this way of minimizing mutual information may result in an unexpected outcome: the mutual information may be minimized via making $\Uin$ contain as little information as possible. To better illustrate the possible outcome, we design $q^u_{1}$ as a linear function which is well trained in Step 1 to ensure Equation (\ref{eq:vCLUB_upper}) is an upper bound of $\I(\Uin;\Uex)$.
Figure \ref{fig:sinDis_problem} shows how the unexpected result may occur. 
%In this situation, when $\Uin$ is inputted into $q^u_{1}$, it can predict the corresponding $\Uex$ well. 
In Step 2, $\Uex, \Uin$ will be trained to minimize Equation (\ref{eq:vCLUB_upper}). To achieve this goal, it ensures $q^u_{1}$ cannot predict $\Uex$ given the corresponding $\Uin$ from the joint distribution (the first term of Equation (\ref{eq:vCLUB_upper})), and at the same time ensures the output of $q^u_{1}$ is similar to the other $\Uex$'s from the marginal distribution (the second term of Equation (\ref{eq:vCLUB_upper})). 

From $\Uin$ perspective (blue circles), the goal can be achieved by pushing the $\Uin$ to move from its original position (optimizing the first term of Equation (\ref{eq:vCLUB_upper})), and move towards the mean of the other $\Uin$'s (optimizing the second term of Equation (\ref{eq:vCLUB_upper})).
From $\Uex$ perspective (red circles), the goal can be achieved by pushing the $\Uex$ away from its original position (optimizing the first term of Equation (\ref{eq:vCLUB_upper})) and the mean of the other $\Uex$'s (optimizing the second term of Equation (\ref{eq:vCLUB_upper})).

This clusters all the $\Uin$'s together, making $\Uin$'s contain less information, while all the $\Uex$'s try to split away from each other, making $\Uex$'s contain more information. The mutual information minimization procedure is like ``transferring'' the information from $\Uin$'s to $\Uex$'s, which is not what we expect.
%It will result in a worse disentanglement than our \textit{BiDis} disentangling component in Figure \ref{fig:ie_vis_all} since 
\textit{BiDis}, however, is a symmetric disentangling method on $\Uin$'s and $\Uex$'s that does not result in this issue.
This may be why \textit{vCLUB} performs worse and is less robust than our proposed symmetrical disentangling component. 

\section{Experiments}
\iffalse
We conduct experiments to demonstrate the effectiveness of the proposed methods.
\begin{itemize}[leftmargin=*]
    \item \textit{\textbf{RQ1:}} How does the proposed IEDR framework perform compared to the state-of-the-art recommendation methods? %in particular, \textcolor{red}{the context-aware and the features-aware ones}?
    % \item \textbf{RQ2:} 
    \item \textit{\textbf{RQ2:}} Can we achieve the intrinsic-extrinsic disentanglement? %using the self-supervised contrastive module and the disentanglement module?
    % \item \textit{\textbf{RQ4:}} Can the proposed IEDR guarantee model generalizability?
        \item \textit{\textbf{RQ3:}} %What is the role of the model components, including 
    % Is it necessary to use two separate embedding vectors to represent a user (or an item)?
    What is the role of the different model components?
\end{itemize}
\fi

We conduct extensive experiments to demonstrate the effectiveness of our model. In this section, we focus on 1) the recommendation performance of IEDR compared to the state-of-the-art methods; 2) the effectiveness of each component in IEDR; and 3) the ability to disentangle intrinsic and extrinsic factors of IEDR.

\subsection{Experimental Setting}
\label{appx:exp_setting}

This section demonstrates the detailed experimental setting to evaluate our method, including the datasets, the baseline methods, and the implementation details.

\subsubsection{Datasets} We evaluate our models in two scenarios with various contexts: a mobile app recommendation and a restaurant recommendation. 
In the mobile app recommendation, we use the Frappe \cite{baltrunas2015frappe} dataset that records mobile app usage logs. 
Each data sample logs users' app usage in a certain context (e.g., weather, time, location). 
%In addition, some relevant properties of the apps are also captured (e.g., category, developer).
In the restaurant recommendation, we use the Yelp dataset \cite{wu2022graph}. Each data sample records users' reviews of local restaurants. 
Due to the fact that a user usually goes to restaurants in the same city, geographic isolation appears in the dataset. Therefore, we select the records in New York City.
We regard each record as a data sample that the user has been to the restaurant. We leverage the user/item features and context features (e.g., day of the week) to predict whether a user will go to a given restaurant in a specific context. 
We also evaluate our model on two Amazon datasets (Movies and CDs) \cite{mcauley2015image}, which have been used in sequential recommendation tasks \cite{yu2019adaptive}. The datasets contain user-item interactions with timestamps. 
For the sequential recommendation, we use the same IEDR model structure as that for the Frappe and Yelp datasets, but modify the data input to fit our model. More specifically, we do not directly learn behavior sequences, but consider each behavior as a data sample with time context information. 
That is, we consider the bucketed timestamp of each user behavior as a time context (we consider one month as a categorized time context). Therefore, behaviors in the same time interval have the same time context, indicating that these behaviors share some similar short-term (extrinsic) interests (e.g., item popularity). 
%For IEDR, we transfer the timestamp to categorized time contexts by bucketing the timestamp with the interval of one month (i.e., the timestamps in the same month are considered as the same context). The datasets also have an item category information as item features.
%\blue{
Note that our experiments are to evaluate our key motivation: learning better intrinsic/extrinsic factor representations. Therefore, our chosen four datasets have high-quality user feedback (e.g., review/comment-based), which is more suitable than other datasets that are larger but less accurate (e.g., click-through-based).
%}

For each dataset, the users that have more than 5 records (Frappe and Yelp) or more than 20 records (Movies and CDs) are chosen. We use the last and the second last record of each user for testing validation, respectively. The rest are for training.
Each of these data samples is considered a positive sample ($y=1$). 
For each positive data sample in the training set, we randomly sample 2 items (but keep the user and contexts) as negative samples ($y=0$), meaning the user did not select the 2 items in that context. For each test/validation data sample, we randomly choose 99 items as negative samples to ensure a more robust evaluation. The statistics of the datasets are shown in Table \ref{tab:datasets}.

\begin{table*}[ht]
\centering
%\small
\caption{Dataset statistics. ``Count'' refers to the number of users/items, and ``Features'' represents the number of different features (for User and Item, the number of features excludes the user/item IDs).}
\label{tab:datasets}
\begin{tabular}{lcccccccc}
\toprule
\textbf{Datasets} & \multicolumn{3}{c}{\textbf{Data Samples}} & \multicolumn{2}{c}{\textbf{User}} & \multicolumn{2}{c}{\textbf{Item}} & \textbf{Context} \\
\cmidrule(lr{0.5em}){2-4}\cmidrule(lr{0.5em}){5-6}\cmidrule(lr{0.5em}){7-8}\cmidrule(lr{0.5em}){9-9}
 & Train & Valid & Test & Count & Features & Count & Features & Features \\
\midrule
Frappe & 282,426 & 69,500 & 69,500 & 695 & 0 & 4,082 & 2,892 & 318\\
 Yelp & 518,208 & 633,600 & 633,600 & 6,336 & 24 & 12,902 & 66 & 13,034\\
\cmidrule(lr{0.5em}){1-9}
Movies & 2,305,362 & 39,663 & 1,322,100 & 13,221 & 0 & 49,189 & 161 & 193 \\
CDs & 879,030 & 16,392 & 546,400 & 5,464 & 0 & 16,184 & 209 & 195  \\
\bottomrule
\end{tabular}
\end{table*}

\begin{table*}[t]
\small
\centering
%\vskip -0.
\caption{Comparing the prediction performance (in percentage) with the baselines. The best-performing results are in bold and the second best are underlined. The \textit{Improv} and \textit{p-value} rows show the relative improvements and the statistical significance of IEDR over the best-performed baselines, respectively.} %N@k refers to NDCG@k and H@k refers to HR@k.}
\label{tab:performance}
\resizebox{0.98\textwidth}{!}{%
\begin{tabular}{lcccccccccc}
\toprule
 & \multicolumn{5}{c}{\textbf{Frappe}} & \multicolumn{5}{c}{\textbf{Yelp}} \\
\cmidrule(lr{0.5em}){2-6}\cmidrule(lr{0.5em}){7-11}
 & NDCG@5 & NDCG@10 & Recall@5 & Recall@10 & AUC & NDCG@5 & NDCG@10 & Recall@5 & Recall@10 & AUC\\
\midrule
AFM        & 63.52  & 67.44  & 77.84  & 84.71  & 93.18  & 42.79  & 47.17  & 58.69  & 72.21 & 91.96\\
NFM        & 68.30  & 70.73  & 83.00  & \underline{90.40}  & 95.86  & 45.99  & 50.33  & 61.90  & 75.27 & 93.32  \\
AutoInt    & \underline{69.45}  & 71.41  & \underline{84.04}  & 90.10  & 95.83  & 46.61  & 50.80  & \underline{63.72}  & \underline{76.55}  & 93.82  \\	
DeepFM     & 69.20  & 71.28  & 82.70  & 89.50  & \underline{96.09}  & 44.20  & 48.50  & 60.26  & 73.55 & 93.26   \\	
WDL        & 68.02  & 70.33  & 81.70  & 88.90  & 95.96  & 45.47  & 49.71  & 61.90  & 74.89 & 93.41    \\
%DCN        & 68.78  & 71.20  & 82.59  & 90.07  & 95.48  & 43.91  & 48.31  & 60.20  & 73.77  & 93.40  \\	
DCNv2      & 68.15  & 70.34  & 82.15  & 89.91  & 95.25  & 43.41  & 48.26  & 60.97  & 74.88  & 93.66  \\	
CL4CTR      & 68.36  & 70.51  & 82.23  & 89.82  & 95.48  & 45.05  & 49.80  & 63.24  & 76.29  & 93.54  \\
EulerNet & 68.87  & 70.68  & 83.30  & 90.36  & 95.88  & 44.81  & 49.54  & 63.33  & 76.08  & 93.47  \\
IFM        & 66.91  & 69.13  & 80.90  & 87.60  & 95.32  & 46.74  & 50.86  & 63.04  & 75.69  & \underline{93.83}  \\
SIGN      & 69.38  & \underline{71.49} & 83.91 & 90.37 & 95.92  & \underline{46.80} & \underline{50.94} & 63.68 & 76.41 & 93.67\\		
%\cmidrule{2-6}\cmidrule{8-12}
DisRec    & 56.81 & 60.07 & 67.42 & 76.29 & 85.51 & 34.82  & 37.90  & 48.29  & 63.17 & 84.01   \\	
DGCF       & 58.40 & 61.44 & 69.05 & 77.53 & 86.13 & 36.35  & 39.06  & 50.05  &  64.62 & 85.29  \\	
\cmidrule(lr{0.5em}){2-6}\cmidrule(lr{0.5em}){7-11}
IEDR & \textbf{72.40 } & \textbf{74.11 } & \textbf{85.94 } & \textbf{91.25 } & \textbf{96.34} & \textbf{48.68 } & \textbf{53.05 } & \textbf{65.23 } & \textbf{78.29 } & \textbf{94.22 } \\ 		
\cmidrule(lr{0.5em}){2-6}\cmidrule(lr{0.5em}){7-11}
\textit{Improv}  & 4.24\% & 3.66\% & 2.26\% & 0.94\% &  0.26\% & 4.01\% & 4.14\% & 2.38\% & 2.28\% & 0.42\% \\
\textit{p-value} & 0.25\% & 0.25\% & 0.25\% & 0.83\% &  3.72\% & 0.25\% & 0.25\% & 0.25\% & 0.25\% & 2.34\% \\
\bottomrule
\end{tabular}
}
%\vskip -0.05in
\end{table*}

%\textcolor{red}{Is it possible to add some data investigation to reveal the necessity of intrinsic-extrinsic disentanglement?}
\subsubsection{Baseline methods}
IEDR models the feature interactions of users, items, and contexts. Therefore, we compare our model with competitive feature interaction-based recommendation methods. 
The methods include 
attentional factorization machine (AFM) \cite{xiao2017attentional}, 
neural factorization machine (NFM) \cite{he2017neural},
self-attention-based feature interaction model (AutoInt) \cite{song2019autoint},
deep factorization machine (DeepFM) \cite{guo2017deepfm}, 
wide \& deep model (WDL) \cite{cheng2016wide},
improved deep \& cross network (DCNv2) \cite{wang2021dcn}, input-aware factorization machine (IFM) \cite{yu2019input}, model-agnostic contrastive learning for CTR (CL4CTR) \cite{wang2023cl4ctr}, and adaptive learning via Euler's formula (EulerNet) \cite{tian2023eulernet}.
We implement these methods using the DeepCTR package or their officially released code.
%with the Apache License 2.0\footnote{\url{https://deepctr-doc.readthedocs.io/en/latest/index.html}}.
%The statistical interaction graph neural network (SIGN) \cite{su2021detecting} is applied based on the released code. %with MIT license.
The above methods model all the factors in a unified representation without considering the factors that affect user behavior.

Meanwhile, we compare IEDR with the methods that learn implicit factors.
%disentangle the user and item information into several representations, with each representation as an implicit factor. 
They are disentangled variational auto-encoder for recommendation (DisRec) \cite{ma2019learning} and disentangled graph collaborative filtering (DGCF) \cite{wang2020disentangled}.
We implement these methods based on their released code.
Note that since DisRec and DGCF models do not consider any feature, their task is to simply predict whether a user will select an item. IERD and other baseline models, however, consider the user-item interactions in specific contexts (a user's behavior in selecting an item may be different in different contexts). For DisRec and DGCF, to prevent the test data samples from appearing in the training set, we remove the data samples from the training set that appear in the test set (with different contexts in other models).
For a fair comparison, we set the factor number to 4 for DisRec and DGCF.
For sequential recommendation baselines, we compare our model with the models that consider LS-term interests. They are
session-based recommender systems with recurrent neural networks (GRU4Rec) \cite{hidasi2016session},
Short-term and Long-term preference Integrated Recommender system (SLI-Rec) \cite{yu2019adaptive},
and Contrastive learning framework of Long and Short-term interests for Recommendation (CLSR) \cite{zheng2022disentangling}.
We use the same MLP structure for feature interaction modeling and the same embedding size for features as our IEDR model.

\subsubsection{Implementation details}
In IEDR, all the MLPs have the same hidden structure: one hidden layer of 128 dimensions and a ReLU activation after that. 
The input and output sizes of MLPs vary based on their needs. 
We set the embedding dimension to 32 for all the features. 
$f_{ie}$ is an MLP that outputs a 64-dimension vector, with the first 32 dimensions being the intrinsic factor representation and the last 32 dimensions being the extrinsic factor representation. 
For the second (dropout-based) negative context-generating method in the context-invariant contrastive learning component, the dropout rate is set to 0.5.
%In the contrastive learning component, the two augmentation methods to generate positive pairs (either randomly sample from the batch or generate new context representation using dropout) are randomly chosen for each data sample. The dropout rate is set to 0.5 for the augmentation.
The number of negative pairs for contrastive learning is 40 for each data sample (note that the actual negative pairs will be doubled since both $(\Uin)_{ii}$ and $(\Uin)_{ij}$ will generate 40 negative pairs). The temperature $\tau$ is set to 0.5.
In the disentangling component, $q_{1}$ and $q_{2}$ are MLPs that output vectors that have the same dimension of intrinsic/extrinsic factor representations. The number of negative samples of the bidirectional vCLUB-based method is 5 for each direction. 
%We use a mean-square error based loss to optimize $\mathcal{L}_{bi-appr}(u,c)$ and $\mathcal{L}_{Dis}(u, c)$.
We set $\lambda_1$ to 0.1 for the Frappe dataset and 0.01 for the Yelp dataset, and set $\lambda_2$ to 0.1 for both datasets.
The $\lambda_1$ and $\lambda_2$ are both 0.01 for the Movies and the CDs datasets.
%We run all the experiments on a machine equipped with a CPU: Intel(R) Xeon(R) Platinum 8163 CPU @ 2.50GHz, and a GPU: Nvidia Tesla v100 GPU.

The model structure of IEDR and its variations used in the experiments are detailed in Table \ref{tab:implement_RP} and Table \ref{tab:implement_cied}. Note that the component structures of variations are the same as the IEDR if not specified.

\subsection{Overall Performance}
\label{sec:exp_overall}

We evaluate the recommendation performance of our model, by comparing it with various baselines in two scenarios. 
In the first scenario, we learn intrinsic and extrinsic factors from various contexts. 
In the second scenario, we learn the factors from a specific (time) context and compare our model with sequential recommendation baselines.
We use three common evaluation metrics for recommender systems: NDCG@$k$, Recall@$k$ with $k$ being 5 and 10, and AUC.

\subsubsection{Factor Learning from Specific Context}

\begin{table}[t]
\caption{ Comparing the performance of IEDR and the baselines on time context-specific scenarios.}
\label{tab:seq_performance}
%\begin{wraptable}{r}{6.9cm}
%\small
\centering
%\vskip -0.1in
\begin{tabular}{lcccc}
\toprule
 & \multicolumn{2}{c}{\textbf{Movies}} & \multicolumn{2}{c}{\textbf{CDs}} \\
\cmidrule(lr{0.5em}){2-3}\cmidrule(lr{0.5em}){4-5}
& NDCG@10 & AUC & NDCG@10  & AUC \\
\midrule
GRU4Rec & 25.18 & 77.11  & 19.41 & 78.86 \\
SLI-Rec & 26.85 & 78.69  & 20.27 & 79.37 \\
CLSR & 26.98 & 80.02 & 21.07 & 80.42 \\
AutoMLP  & 26.73 & 79.91 &  20.32 & 79.60 \\
BERT4Rec  & 25.27 & 78.24 & 19.53 & 79.13 \\
SASRec & 26.28 & 79.49 &  20.67 & 79.71 \\
S3Rec  & \textbf{27.04} & 80.11 & 21.16 & 80.09 \\
TiSASRec & 26.84 & 79.93 & \textbf{21.25} & 80.18 \\
%\cmidrule(lr{0.5em}){2-3}\cmidrule(lr{0.5em}){4-5}
\midrule
AutoInt	&22.27	&77.78	&18.25  &77.65\\
DeepFM	&23.13	&78.50	&19.18  &78.26\\
SIGN	&23.58	&78.82	&19.97 	&78.95\\
\midrule
%\cmidrule(lr{0.5em}){2-3}\cmidrule(lr{0.5em}){4-5}
%IEDR & \textbf{96.34} & \textbf{72.40 } & \textbf{74.11 } & \textbf{85.94 } & \textbf{91.25 } & \textbf{94.22 } & \textbf{48.68 } & \textbf{53.05 } & \textbf{65.23 } & \textbf{78.29 }\\ 		
IEDR & 26.68 & \textbf{80.14}  & 20.95 & \textbf{80.34}\\
%\cmidrule(lr{0.5em}){2-3}\cmidrule(lr{0.5em}){4-5}
%\textit{Improv}  &  0.26\% & 4.24\% & 3.66\% & 2.26\% & 0.94\% & 0.42\% & 4.01\% & 4.14\% & 2.38\% & 2.28\% \\
%\textit{p-value} &  3.72\% & 0.25\% & 0.25\% & 0.25\% & 0.83\% & 2.34\% & 0.25\% & 0.25\% & 0.25\% & 0.25\% \\
\bottomrule
\end{tabular}
%\end{wraptable}
\end{table}
%\vskip -0.20in

\iffalse
\begin{table}[t]
%\begin{wraptable}{r}{6.9cm}
%\small
\centering
%\vskip -0.1in
\begin{tabular}{lcccc}
\toprule
 & \multicolumn{2}{c}{\textbf{Movies}} & \multicolumn{2}{c}{\textbf{CDs}} \\
\cmidrule(lr{0.5em}){2-3}\cmidrule(lr{0.5em}){4-5}
 & AUC & NDCG@10 & AUC & NDCG@10 \\
\midrule
GRU4Rec & 77.11  & 25.18 & 78.86 & 19.41\\
SLI-Rec & 78.69 & 26.85 & 79.37 & 20.27 \\
CLSR & 80.02 & 26.98 & 80.42 & 21.07 \\
%\cmidrule(lr{0.5em}){2-3}\cmidrule(lr{0.5em}){4-5}
\midrule
AutoInt	&77.78	&22.27	&77.65	&18.25 \\
DeepFM	&78.50	&23.13	&78.26	&19.18 \\
SIGN	&78.82	&23.58	&78.95	&19.97 \\
\midrule
%\cmidrule(lr{0.5em}){2-3}\cmidrule(lr{0.5em}){4-5}
%IEDR & \textbf{96.34} & \textbf{72.40 } & \textbf{74.11 } & \textbf{85.94 } & \textbf{91.25 } & \textbf{94.22 } & \textbf{48.68 } & \textbf{53.05 } & \textbf{65.23 } & \textbf{78.29 }\\ 		
IEDR & 80.14 & 26.68 & 80.34 & 20.95 \\
%\cmidrule(lr{0.5em}){2-3}\cmidrule(lr{0.5em}){4-5}
%\textit{Improv}  &  0.26\% & 4.24\% & 3.66\% & 2.26\% & 0.94\% & 0.42\% & 4.01\% & 4.14\% & 2.38\% & 2.28\% \\
%\textit{p-value} &  3.72\% & 0.25\% & 0.25\% & 0.25\% & 0.83\% & 2.34\% & 0.25\% & 0.25\% & 0.25\% & 0.25\% \\
\bottomrule
\end{tabular}
\caption{ Comparing the performance of IEDR and the baselines on time context-specific scenario.}
\label{tab:seq_performance}
%\end{wraptable}
\end{table}
\fi

We then evaluate IEDR on two Amazon datasets (Movies and CDs) \cite{mcauley2015image} that contain only the time context.
We compare with the state-of-the-art sequential recommendation baselines
GRU4Rec \cite{hidasi2016session}, LSI-Rec \cite{yu2019adaptive}, CLSR \cite{zheng2022disentangling} and AutoMLP \cite{li2023automlp}, that learn long-short term interests from the item sequences ordered by the time (discussed in Section \ref{sec:relativeWork_factorDisentanglement}).
Also, we compare with state-of-the-art general sequential recommendation baselines, BERT4Rec \cite{sun2019bert4rec}, SASRec \cite{kang2018self}, S3Rec \cite{zhou2020s3}, TiSASRec \cite{wang2020time}.
In IEDR, we use the same model structure as that for the Frappe and Yelp datasets, but modify the data input to fit our model. More specifically, without directly learning behavior sequences, IEDR considers each behavior as a data sample with time context information, where the time context is the bucketed timestamp of each user behavior (one month as a categorized time context). 
We also run the best-performing baselines from Table \ref{tab:performance} on the Amazon datasets.
The experimental results are reported in Table \ref{tab:seq_performance}.

From these results, we can see that our model achieves competitive accuracy compared to the sequential recommendation baselines.
This proves the ability of our model to achieve state-of-the-art recommendation accuracy in the context-specific scenario, even compared with the models designed for the context.
%Meanwhile, the advantage of our IEDR is that it is more versatile, and can be applied to various contexts as well.
Moreover, our IEDR is more versatile and can be applied to various contexts.
Finally, the feature interaction-based baselines do not disentangle intrinsic and extrinsic factors. Therefore, they perform worse than our models and sequential recommendation baselines on the Amazon datasets.

\subsection{Effectiveness of Our Model's Components}\label{sec:RQ3}
\label{sec:ablation}
\iffalse
\begin{enumerate}
    \item Graph-based input / feature-based concat input / FM input
    \item w/wo CL module, CL methods [mask/sampling]
    \item w/wo disentanglement module \\ Unidirectional \textit{v.s.} bidirectional disentanglement (average \& variance)
\end{enumerate}
\fi
This section evaluates the components of IEDR. We only demonstrate the results in NDCG@10 since metrics show similar trends. 
%Other evaluation metrics show similar trends and we list them in Appendix.

\subsubsection{Ablation Study of Contrastive Intrinsic-Extrinsic Disentangling Module}
\iffalse
The contrastive intrinsic-extrinsic disentangling (CIED) module contains a context-invariant contrastive learning component and a disentangling component. In this section, we conduct an ablation study to show the impact of these components.
We run our model in three variations: 1) without the contrastive learning component (\textit{noCL}); 2) without the disentangling component (\textit{noDis}); 3) without the contrastive learning and disentangling components (\textit{noCIED}), i.e., the CIED module is not applied. 
Figure \ref{fig:ablation} compares our IEDR model with these three variants. 
The inferior performance of \textit{noCIED} compared with our full model \textit{IEDR} demonstrates the importance of learning intrinsic and extrinsic factors for accurate recommendation prediction.
\textit{noCL} can be regarded as performing implicit factor learning. 
The inferior performance of \textit{noCL} 
indicates that explicit intrinsic and extrinsic factor learning is superior to implicit factor learning methods (as in the factor disentangling baselines) in analyzing user behavior.
\textit{noDis} learns intrinsic factors but does not guarantee extrinsic factor learning. Therefore, it also obtains worse results than CIED.  
Employing both components achieves better performance than learning with only one. This is because either component cannot individually learn intrinsic and extrinsic factors successfully, highlighting the importance of learning the two factors for an accurate recommendation.
\fi

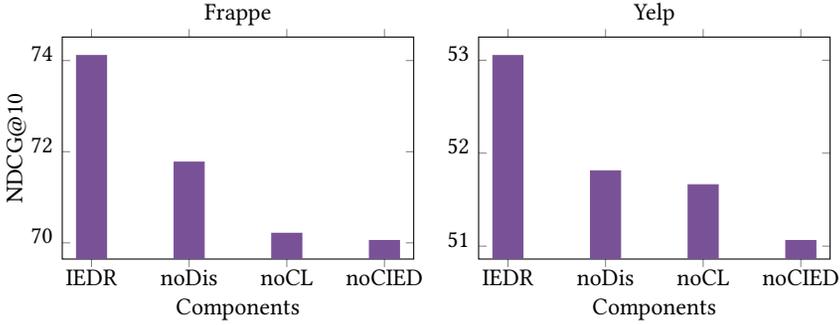
\begin{figure}[t]
    \centering
        \begin{tikzpicture}
            \begin{axis}[barstyle,width=0.45\textwidth, title={Frappe}, ylabel={NDCG@10}, xlabel={Components},bar width=4mm, symbolic x coords = {IEDR, noDis, noCL, noCIED}, legend style={draw=none, at={(0.4,0.8)}, anchor=west, nodes={scale=0.9, transform shape}, legend image post style={scale=0.5}}]
            \addplot [color=mycolor2, fill=mycolor2] coordinates { 
            (IEDR, 74.11)
            (noDis, 71.77)
            (noCL, 70.21)
            (noCIED, 70.05)
            };
            \end{axis}
            \end{tikzpicture}
            \begin{tikzpicture}
            \begin{axis}[barstyle,width=0.45\textwidth, title={Yelp}, xlabel={Components},bar width=4mm, symbolic x coords = {IEDR, noDis, noCL, noCIED}, legend style={draw=none, at={(0.4,0.8)}, anchor=west, nodes={scale=0.9, transform shape}, legend image post style={scale=0.5}}]
            \addplot [color=mycolor2, fill=mycolor2] coordinates { 
            (IEDR, 53.05)
            (noDis, 51.81)
            (noCL, 51.66)
            (noCIED, 51.06)
            };
            \end{axis}
        \end{tikzpicture}
        \caption{Ablation studies results with different component(s) removed.}
        \Description{Ablation studies results with different component(s) removed.}
        \label{fig:ablation}
\end{figure}

To evaluate the contribution of the Contrastive Intrinsic-Extrinsic Disentangling (CIED) module, we compare IEDR against three variants: \textit{noDis} (removes the disentanglement component), \textit{noCL} (removes the context-invariant contrastive learning component), and \textit{noCIED} (removes both components). The experiments are conducted on the Frappe and Yelp datasets, and the results are presented in Figure \ref{fig:ablation}.
The results highlight the synergistic contribution of the two components in IEDR. 1) \textit{IEDR} achieves the best performance on both datasets (74.11 on Frappe and 53.05 on Yelp), with improvements over \textit{noCIED} of 4.06 points on Frappe and 2.99 points on Yelp, exceeding the combined individual improvements of \textit{noDis} and \textit{noCL}. This indicates a cumulative effect, where the disentanglement component and CICL reinforce each other, ensuring stable intrinsic factors and effective separation of extrinsic factors. 2) The small improvement of \textit{noCL} over \textit{noCIED} on Frappe (0.16 points) highlights the limitations of relying solely on implicit factor disentanglement, particularly in datasets dominated by context features. These findings emphasize the importance of explicit factor learning through CICL, which ensures robust disentanglement and overall performance gains.

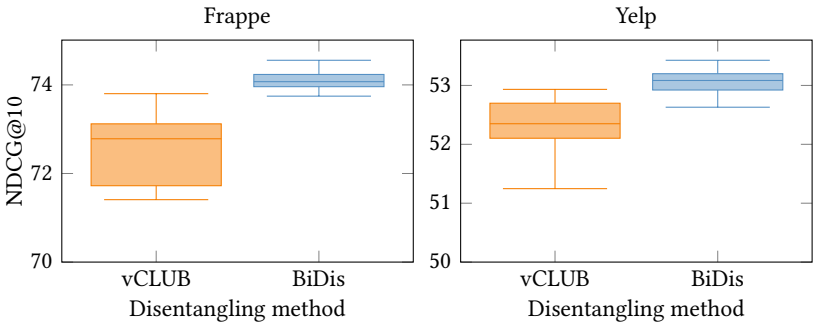
\begin{figure}[t]
    \centering
    \begin{tikzpicture}
      \begin{axis}
        [boxstyle,width=0.45\textwidth,ylabel={NDCG@10},xlabel={Disentangling method}, xticklabels={vCLUB, BiDis}, ymin=70, title={Frappe}]
        \addplot+[
        fill,fill opacity=0.5,
        boxplot prepared={
          median=72.784,
          upper quartile=73.121,
          lower quartile=71.724,
          upper whisker=73.802,
          lower whisker=71.409
        },
        ] coordinates {};
        \addplot+[
        fill,fill opacity=0.5,
        boxplot prepared={
          median=74.073,
          upper quartile=74.236,
          lower quartile=73.960,
          upper whisker=74.556,
          lower whisker=73.746
        },
        ] coordinates {};
      \end{axis}
    \end{tikzpicture}
    \begin{tikzpicture}
      \begin{axis}
        [boxstyle,width=0.45\textwidth,xlabel={Disentangling method}, xticklabels={vCLUB, BiDis}, ymin=50,title={Yelp}]
        \addplot+[
        fill,fill opacity=0.5,
        boxplot prepared={
          median=52.3495,
          upper quartile=52.698,
          lower quartile=52.103,
          upper whisker=52.933,
          lower whisker=51.247
        },
        ] coordinates {};
        \addplot+[
        fill,fill opacity=0.5,
        boxplot prepared={
          median=53.084,
          upper quartile=53.199,
          lower quartile=52.922,
          upper whisker=53.427,
          lower whisker=52.630
        },
        ] coordinates {};
      \end{axis}
    \end{tikzpicture}
    %\vskip -0.05in
    \caption{The performance and variance statistics of vCLUB and BiDis.}
    \Description{The performance and variance statistics of vCLUB and BiDis.}
    \label{fig:bi_vs_sing_perform}
\end{figure}
%\vskip -0.2in

\subsubsection{Disentangling Component Evaluation}
\label{sec:vCLUBvsbiDis}
%1. vCLUB vs bi-vCLUB (box plot) vs no disentangle
%2. *Possible disentanglement visualization
We propose a bidirectional vCLUB-based disentangling method (\textit{BiDis}) to disentangle the intrinsic and extrinsic factors. In this section, we compare our \textit{BiDis} method with the original vCLUB method (\textit{vCLUB}) \cite{cheng2020club} in model performance. 
The results in Figure \ref{fig:bi_vs_sing_perform} highlight the superiority of our \textit{BiDis} method over \textit{vCLUB} in both performance and robustness. \textit{BiDis} leverages bidirectional mutual information minimization, ensuring a more thorough and balanced disentanglement of intrinsic and extrinsic factors, as discussed in Section \ref{sec:method_biDis}. This bidirectional approach avoids the instability and noise issues associated with vCLUB’s asymmetric optimization, resulting in more robust and consistent performance across datasets. Additionally, the visualization in Section \ref{sec:more_vis} further demonstrates that \textit{BiDis} produces clearer and more distinct factor separation, underscoring its effectiveness in real-world recommendation scenarios.

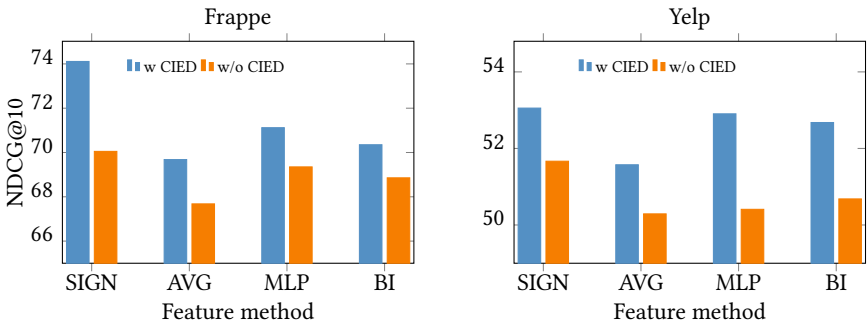
\begin{figure}[b]
    \centering
    \begin{tikzpicture}
        \begin{axis}[barstyle, width=0.45\textwidth,title={Frappe}, ylabel={NDCG@10}, xlabel={Feature method},bar width=3mm, symbolic x coords = {SIGN, AVG, MLP, BI}, legend style={draw=none, at={(0.16,0.88)}, anchor=west, nodes={scale=0.7, transform shape}, legend image post style={scale=0.5}}, ymin=65]
        \addplot [color=mycolor3, fill=mycolor3] coordinates { 
        (SIGN, 74.11 )
        (AVG, 69.68)
        (MLP, 71.12 )
        (BI, 70.35)
        };
        \addplot[color=mycolor1, fill=mycolor1] coordinates { 
        (SIGN, 70.05)
        (AVG, 67.68)
        (MLP, 69.35 )
        (BI, 68.856)
        };
        \legend{w CIED, w/o CIED}
        \end{axis}
        \end{tikzpicture}
        \hskip 2em
        \begin{tikzpicture}
        \begin{axis}[barstyle,width=0.45\textwidth, title={Yelp}, xlabel={Feature method},bar width=3mm, symbolic x coords = {SIGN, AVG, MLP, BI}, legend style={draw=none, at={(0.16,0.88)}, anchor=west, nodes={scale=0.7, transform shape}, legend image post style={scale=0.5}}, ymin=49, ymax=54.8]
        \addplot [color=mycolor3, fill=mycolor3] coordinates { 
        (SIGN, 53.054 )
        (AVG, 51.573)
        (MLP, 52.904 )
        (BI, 52.677)
        };
        \addplot[color=mycolor1, fill=mycolor1] coordinates { 
        (SIGN, 51.666 )
        (AVG, 50.29)
        (MLP, 50.409 )
        (BI, 50.681)
        };
        \legend{w CIED, w/o CIED}
        \end{axis}
    \end{tikzpicture}
    %\vskip -0.15in
    \caption{Model performance when equipped with different feature modeling methods.}
    \label{fig:feature_model}
\end{figure}

\subsubsection{Other Feature Modeling Methods}
\label{appx:exp_otherFM}
%Graph-based, meanfeature, MLP, FM (Bar chart)
In the RP module, although we use a SIGN-based method \cite{su2021detecting} to learn user, item, and context features, the module can use any feature modeling method. Here, we use other methods to evaluate whether our model still performs well. 
Specifically, we run our model with the other three variations using different feature modeling methods: 1) averaging feature embeddings (\textit{MEAN}); 2) adding an MLP on top of the averaged feature embedding (\textit{MLP}); and 3) modeling and aggregating feature interactions through a Bi-interaction layer proposed in \cite{he2017neural} (\textit{BI}). 
The results are shown in Figure \ref{fig:feature_model}. We report the results of each variation with and without the CIED module. 
From this figure, we can see that when equipped with the CIED module, all feature modeling methods perform better than those without the module. It shows that our proposed CIED module can learn intrinsic and extrinsic factors for more accurate recommendations when different feature modeling methods are applied.
Meanwhile, the feature modeling methods can impact the performance. \textit{MEAN} is just a linear aggregation of features, resulting in the worst performance. Both \textit{MLP} and \textit{BI} have better feature modeling ability and hence have better performance than \textit{MEAN}. The SIGN-based feature modeling (\textit{SIGN}) is the state-of-the-art feature interaction modeling method and archives the best performance.

\subsection{Comparing the Impact of Different Contrastive Learning Variations}
\label{appx:exp_cl_variant}
\begin{table}[t]
\centering
\caption{ Comparing the performance of $\text{IEDR}_{sp}$ with different dropout rates (for \textit{NegGen2}).}
\label{tab:intrinsic_only_user}
\begin{tabular}{lcc}
\toprule
 & Frappe & Yelp \\
\midrule
$\text{IEDR}_{sp}$, p=0.1             & 70.68  & 52.03  \\
$\text{IEDR}_{sp}$, p=0.5             & 68.25  & 51.49  \\
$\text{IEDR}_{sp}$, p=0.1, noDis       & 70.56  & 52.02  \\
$\text{IEDR}_{sp}$, p=0.1, noCL        & 70.31  & 51.62  \\
$\text{IEDR}_{sp}$, p=0.1, noCIED       & 70.16  & 51.10  \\
\midrule
IEDR & \textbf{74.11} & \textbf{53.05} \\ 		
\bottomrule
\end{tabular}
\end{table}

To learn intrinsic factors, we propose a context-invariant contrastive learning method. However, directly generating intrinsic factor representations through user information seems to be a more direct way, i.e., $\Uin=f_{ie}^{u}(\urep)$. 
However, we argue that the intrinsic factors learned this way could not guarantee the effectiveness of intrinsic factor learning. 
This is because the information in the learned intrinsic factor representations can vary with different contexts, since these factors have never been modeled w.r.t. the contexts.

In this section, we empirically evaluate our argument and show that our context-invariant contrastive learning method generates more accurate recommendations. To do so, we design a variation ($\text{IEDR}_{sp}$) by splitting the intrinsic-extrinsic factor generation into two functions: $\Uin=f_{in}^{u}(\urep)$, and $\Uex=f_{ex}^{u}(\urep, \crep)$. Both $f_{in}$ and $f_{ex}$ have the same structure as $f_{ie}$, with the output dimension being a half to ensure the consistency of the factor representation dimension. The contrastive learning component does not consider context information but uses a standard InfoNCE-based contrastive learning for learning robust user/item representations following \cite{yao2021self}.
Table \ref{tab:intrinsic_only_user} illustrates the results of $\text{IEDR}_{sp}$ compared to our model with $\text{IEDR}_{sp}$ using different dropout rates ($p=0.1$ and $p=0.5$) in the contrastive learning component, and different component combinations (\textit{noDis}, \textit{noCL}, \textit{noCIED}).
The experiment demonstrates that our model outperforms the variation in recommendation accuracy. 
This proves that $\text{IEDR}_{sp}$ cannot ensure successful intrinsic factor learning and hence incurs a worse recommendation accuracy.
%It is because $f_{in}(\urep)$ does not guarantee the learned intrinsic factor that fits Definition \ref{def:in_ex_factor}.
Unlike IEDR, $\text{IEDR}_{sp}$ gains better performance with a lower dropout rate. This is because, in $\text{IEDR}_{sp}$, the dropout generates views representing the same user instead of different users, which is consistent with the conclusion in \cite{gao2021simcse}.
%In addition, it is interesting to see that $IEDR_{sp}$ has similar performances no matter quipped with the disentangling component or not. It proves that the splitted factor learning already yield some disentanglement so that the disentangling component does not further perform the performance gain. However, our model learns better intrinsic and extrinsic factor representations so that outperforms $IEDR_{sp}$.

\subsection{Disentanglement Verification}
\iffalse
\begin{enumerate}
    \item Put the visualization of embedding weights with PCA and t-SNE.
    \item Put the case study results [intrinsic-extrinsic matching results - item list \& distribution]
\end{enumerate}
\fi
This section verifies the intrinsic and extrinsic factor disentangling ability of IEDR, including a visualization of the learned intrinsic and extrinsic representations and a case study to show the differences between these factors in users' decision-making. 

\subsubsection{Intrinsic and Extrinsic Representation Visualization}
\label{sec:more_vis}

\begin{figure*}[t]
\centerline{\includegraphics[width=0.6\textwidth]{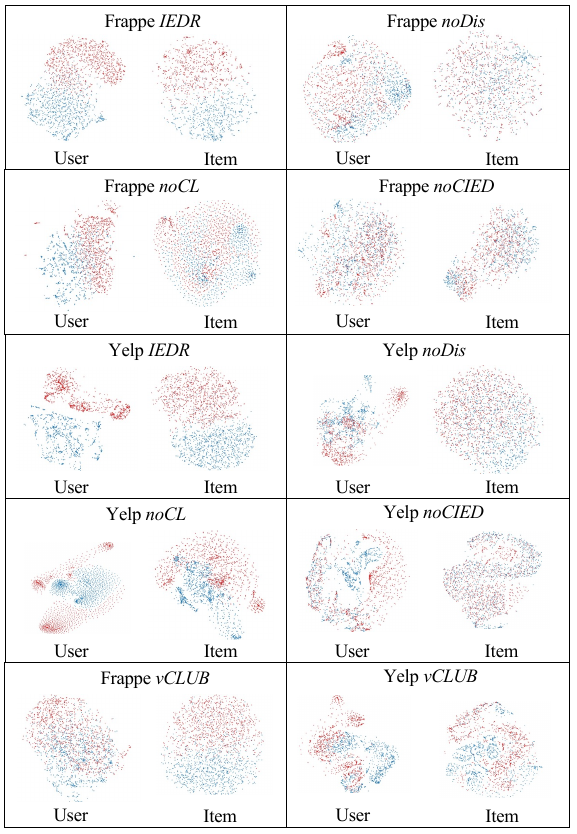}}
%\vskip -0.05in
\caption{ The complete intrinsic-extrinsic disentanglement visualizations in t-SNE. The blue dots are intrinsic representations, and the red dots are extrinsic representations.}
%\vskip -0.1in
\label{fig:ie_vis_all}
\end{figure*}

This section provides intrinsic and extrinsic representation visualizations of our model and three variations: 1) the contrastive learning component is removed (\textit{noCL}); 2) the disentangling component is removed (\textit{noDis}); and 3) the asymmetric disentanglement method (\textit{vCLUB}) is used.
Figure \ref{fig:ie_vis_all} compares these results. We include our main observations below:
\begin{itemize}[leftmargin=*]
    \item The intrinsic and extrinsic factors are perfectly disentangled with our CIED module (\textit{IEDR}).
    \item Without the disentangling component (\textit{noDis}), the intrinsic and extrinsic disentangling procedure may not succeed. This is because there is no restriction on extrinsic representations. Therefore, the extrinsic representations can contain any information, including the information of the intrinsic factor.
    \item \textit{noCL} has worse disentangling performance than \textit{IEDR}, either. This is because the factors disentangled in \textit{noCL} are implicit. The implicit factors only ensure the disentanglement between the factors of the same data sample, but not between the factors of other data samples. For example, some context information may be stored in the intrinsic representation in data sample 1 but be stored in the extrinsic representation in data sample 2.
    \item \textit{noCIED} performs worst among all variations, which is reasonable since it does not distinguish the intrinsic and extrinsic representations.
    \item \textit{vCLUB} performs disentanglement, but is not very stable in some situations. This is consistent with our analysis in Section \ref{appx:vclub_problem}.
\end{itemize}

\iffalse

In Figure \ref{fig:ie_vis_all}, we visualize intrinsic and extrinsic factor representations learned by IEDR.
From the figure, we can observe that:
We see that when the model is equipped with the CIED module (\textit{IEDR}), the factors are well disentangled. 
However, when we do not use the CIED module (\textit{noCIED}), the intrinsic and extrinsic factor representations are mixed together. 
This indicates that these factors cannot be well learned and disentangled without our CIED module.
%We provide further illustrative examples and analysis with different component combinations in Appendix \ref{appx:more_vis}.

%\floatsetup[figure]{style=Boxed}
\begin{figure}[ht]
\centering
\begin{subfigure}{0.22\textwidth}
\includegraphics[width=\textwidth]{figure/user_all1.png}
\caption{\small User \textit{IEDR}.}
\end{subfigure}
%\hfill
\begin{subfigure}{0.22\textwidth}
\includegraphics[width=\textwidth]{figure/item_all1.png}
\caption{\small Item \textit{IEDR}.}
\end{subfigure}
%\hfill
\begin{subfigure}{0.22\textwidth}
\includegraphics[width=\textwidth]{figure/user_noboth.png}
\caption{ User \textit{noCIED}.}
\end{subfigure}
%\hfill
\begin{subfigure}{0.22\textwidth}
\includegraphics[width=\textwidth]{figure/item_noboth1.png}
\caption{\small Item \textit{noCIED}.}
\end{subfigure}
\caption{ Visualization of the learned intrinsic and extrinsic factors with t-SNE for the Frappe dataset. The blue dots are the intrinsic factors, and the red dots are the extrinsic factors.}
%\vskip -0.1in
\label{fig:ie_vis}
\end{figure}
\fi

\subsubsection{Case Study}
\begin{figure}[ht]
\centering
\includegraphics[width=0.40\columnwidth]{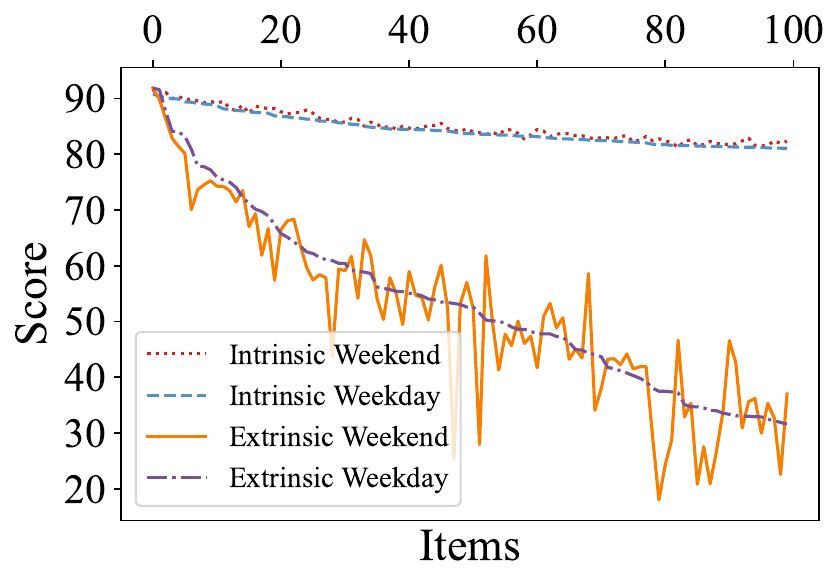}
\caption{ A user's top 100 intrinsic and extrinsic scores in different contexts (Weekend vs. Weekday).}
\label{fig:case_score}
%\vskip -0.10in
\end{figure}

We conducted a case study to analyze the differences between the learned intrinsic and extrinsic factors. 
We randomly choose a user from the Frappe dataset and generate the intrinsic matching scores (the dot product of the user's intrinsic representation and the items' (apps) intrinsic representations) in two different contexts (Weekday and Weekend). The same for the extrinsic matching scores.
We sort the matching scores for the intrinsic and extrinsic factors, respectively, and list the top 100 items.
The results are in Figure \ref{fig:case_score}.
Note that the top 100 items for intrinsic and extrinsic factors are different.
%Note that for the same x-axis value, intrinsic scores are for the same item, and extrinsic scores may be for another item.
According to Figure \ref{fig:case_score}, from weekday to weekend, the extrinsic scores vary a lot, while the intrinsic scores remain invariant. These observations demonstrate that, in different contexts, the user has different intrinsic factors, as well as consistent intrinsic factors.

\begin{table}[t]
\centering
%\scriptsize
\tabcolsep=0.07cm
\caption{ Items (in category) of the highest intrinsic and extrinsic scores for different users in Weekday.}
\label{tab:casestudy}
\begin{tabular}{ccccc}
\toprule
 & \multicolumn{2}{c}{\textbf{User1}} & \multicolumn{2}{c}{\textbf{User2}}\\
\cmidrule(lr{0.5em}){2-3}\cmidrule(lr{0.5em}){4-5}
Rank & Intrinsic & Extrinsic & Intrinsic & Extrinsic \\
\midrule
1 & Photography & Tools & Cards\&Casino & Communication\\
2 & Sports & Communication& Productivity & Tools\\
3 & Health\&Fitness & Media\&Video& Cards\&Casino &News\&Magazines\\
4 & Tools &Personalization& Sports Games &Tools\\
5 & Health\&Fitness &Communication& Brain\&Puzzle &Communication\\
6 & Personalization &Casual& Communication &News\&Magazines\\
7 & Personalization &Music\&Audio& Tools &Personalization\\
8 & Communication &News\&Magazines& Sports &Media\&Video\\
9 & Personalization &Communication& Arcade\&Action &Tools\\
10 & Health\&Fitness &Travel\&Local& Tools &Communication\\
\bottomrule
\end{tabular}
\end{table}

\iffalse
Then, we evaluate how users may have individual intrinsic factors (real interests). Table \ref{tab:casestudy} lists the categories of the items with the 10 highest intrinsic scores for three users, respectively.
From the table, we observe that users have individual intrinsic interests, e.g., \textit{User1} prefers sports and fitness apps while \textit{User2} prefers gaming apps.
Besides, it is intuitive that these apps may more likely be used in personal time instead of working time. The intrinsic factors show users' real interests instead of being dominated by the most used apps (e.g., productivity apps like Gmail and Chrome).
\fi

Then, we show how intrinsic and extrinsic factors may have different impacts on users' choices. Table \ref{tab:casestudy} lists the categories of the items with the 10 highest intrinsic/extrinsic scores for two users, respectively.
we can observe that users have individual intrinsic interests that indicate their real hobbits, e.g., \textit{User1} prefers sports and fitness apps, while \textit{User2} prefers gaming apps.
On the other hand, extrinsic factors give a higher rank to the items based on the contexts (Weekday), e.g., Tools (Google Search) and Communication (Gmail) rank highest in \textit{User1}'s extrinsic scores.

% \subsection{Further analysis}
% \subsubsection{Ablation study on modules}
% \textcolor{red}{Put the results of ablating the essential components of IEDR.}
% \begin{itemize}
%     \item Graph-based input, with v.s. without [Self-supervision+Disentanglement]
%     \item Feature concat input, with v.s. without [Self-supervision+Disentanglement]
% \end{itemize}
% \subsection{Ablation study on hyperparameters}
% Multi-task loss weights: main supervised task + self-supervision + disentanglement 
% \subsubsection{Case study}

\subsection{Different Negative Context Generation Methods}
\label{appx:exp_augs}

\begin{table}[ht]

\caption{Comparing the performance of IEDR using different negative context generating methods (for the contrastive learning component).}
\label{tab:diff_augs}
\centering
\begin{tabular}{lcc}
\toprule
 & Frappe & Yelp \\
\midrule
NegGen1              & 73.01   & 52.49  \\
NegGen2             & 71.50  & 51.82  \\
NegGen1\&2         & \textbf{74.11}  & \textbf{53.05}  \\
\bottomrule
\end{tabular}
\end{table}

\begin{figure}[ht]
\centering
\begin{tikzpicture}
\begin{axis}[linestyle,width=0.45\textwidth, title =Frappe, xlabel={Dropout Rate}, legend columns=1, ymax=85, symbolic x coords={0.1, 0.3, 0.5, 0.7, 0.9}, legend style={draw=none, at={(0.84,0.76)},anchor=east, nodes={scale=0.9, transform shape}}, legend image post style={scale=0.6}]
\addplot[mark=*, color=mycolor1] coordinates {
%(k, Recall@10)
(0.1, 69.63 )
(0.3, 71.41  )
(0.5, 71.50 )
(0.7, 71.43  )
(0.9, 67.94 )
};
\addplot[mark=triangle*,color=mycolor3] coordinates {
%(k, NDCG@10)
(0.1, 70.77 )
(0.3, 73.45 )
(0.5, 74.11 )
(0.7, 74.44 )
(0.9, 71.02 )
};
\legend{\textit{NegGen2}, \textit{NegGen1\&2}}
\end{axis}
\end{tikzpicture}
\begin{tikzpicture}
\begin{axis}[linestyle,width=0.45\textwidth, title =Yelp, xlabel={Dropout Rate}, legend columns=1,ymax=57, symbolic x coords={0.1, 0.3, 0.5, 0.7, 0.9}, legend style={draw=none, at={(0.84,0.76)},anchor=east, nodes={scale=0.9, transform shape}}, legend image post style={scale=0.6}]
\addplot[mark=*, color=mycolor1] coordinates {
%(k, Recall@10)
(0.1, 50.91 )
(0.3, 51.41 )
(0.5, 51.82 )
(0.7, 50.84 )
(0.9, 50.59 )
};
\addplot[mark=triangle*,color=mycolor3] coordinates {
%(k, NDCG@10)
(0.1, 51.64 )
(0.3, 52.41 )
(0.5, 53.05 )
(0.7, 51.86 )
(0.9, 51.30 )
};
\legend{\textit{NegGen2}, \textit{NegGen1\&2}}
\end{axis}
\end{tikzpicture}
\caption{ The performance of different dropout rates for method 2 (\textit{NegGen2}).}
\label{fig:cl_dropoutrate}
\end{figure}
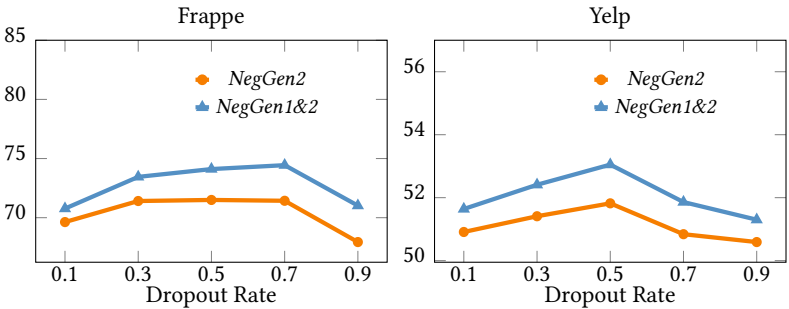

%\subsection{Different negative context generation methods}
%1. Both 1,2 , only 1, only 2 context generation methods
%2. Different droprate (with 1 and without 1 augmentation)
We propose two negative context-generating methods in the contrastive learning component: 1) sample other contexts; 2) use a large dropout rate on the original context. 
%In this section, we perform ablation studies on the two generating methods and evaluate the performance of different dropout rates.
We evaluate the two methods in this section.
Table \ref{tab:diff_augs} shows the results of our model when using only \textit{NegGens1}, only \textit{NegGens2}, and \textit{NegGen1\&2}.
We can see that \textit{NegGen1} results in a better performance than using \textit{NegGen2}. This is because \textit{NegGen1} uses true context representations, which are consistent with what may appear in the test samples. 
Meanwhile, we see that \textit{NegGen1\&2} results in the best performance. This is because \textit{NegGen2} provides more unseen (randomly generated) context representations, which strengthens the generalization ability of our model. 
Next, we evaluate \textit{NegGen2} with different dropout rates in Figure \ref{fig:cl_dropoutrate}.
The best performance can be achieved when the dropout rates range from 0.5 to 0.7. This is consistent with our claim in Section \ref{sec:method_CICL}. The reason is that a small dropout rate (e.g., 0.1) pushes the generated context representation too close to the original one; hence it cannot be considered a different context. However, a relatively large dropout rate (e.g., 0.9) loses too much information; hence, it is no longer a valid context representation. 
In addition, for \textit{NegGen1\&2} of all the dropout rates, the results consistently outperform those that only use \textit{NegGen2}.
%This is consistent with the conclusion draw from Figure \ref{fig:negcontext_gen}.

\subsection{Effectiveness of Model Hyperparameters}
\label{sec:exp_hyper}

We evaluate our model with different hyperparameter settings, including embedding dimensions, number of negative samples, and loss weight values. Below, we summarize our observations.

\subsubsection{Embedding Dimension}

\begin{figure}[ht]
\centering
\begin{tikzpicture}
\begin{axis}[linestyle, width=0.45\textwidth, title =Frappe, xlabel={Embedding dimension}, legend columns=1,symbolic x coords={8, 32, 64, 128, 256}, legend style={draw=none, at={(0.84,0.76)},anchor=east, nodes={scale=0.9, transform shape}}, legend image post style={scale=0.6}]
\addplot[mark=*, color=mycolor2] coordinates {
%(k, Recall@10)
(8, 66.66 )
(32, 74.11 )
(64, 73.646)
(128, 73.6087 )
(256, 72.224)
};
\end{axis}
\end{tikzpicture}
\begin{tikzpicture}
\begin{axis}[linestyle,width=0.45\textwidth, title =Yelp, xlabel={Embedding dimension}, legend columns=1,symbolic x coords={8, 32, 64, 128, 256}, legend style={draw=none, at={(0.84,0.76)},anchor=east, nodes={scale=0.9, transform shape}}, legend image post style={scale=0.6}]
\addplot[mark=*, color=mycolor2] coordinates {
%(k, Recall@10)
(8, 48.59 )
(32, 53.0546)
(64, 52.667)
(128, 53.397)
(256, 53.119)
};
\end{axis}
\end{tikzpicture}
\caption{ The performance of different embedding dimensions.}
\label{fig:diff_dim}
%%\vskip -0.1in
%\vskip -0.1in
\end{figure}
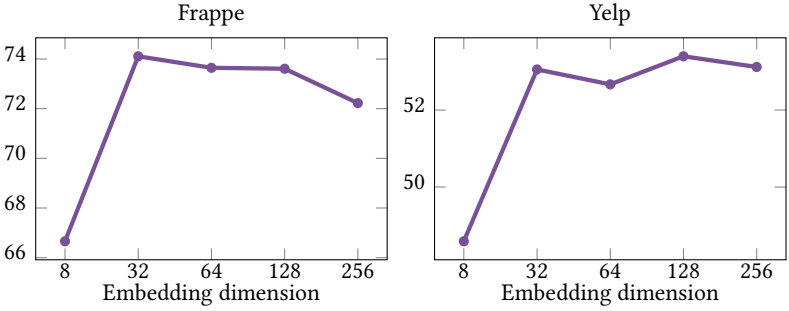

We run our model with different feature embedding dimensions.
The results are in Figure \ref{fig:diff_dim}. 
The embedding dimension poses a trade-off between the expression ability and efficiency.
From the figure, we can see that larger dimensions result in better recommendation accuracy.
However, the improvement is not significant when the dimension is larger than 32. 
A larger dimension may even reduce the performance due to the overfitting problem (e.g., dimension 256 for the Frappe dataset).

\subsubsection{The Number of Negative Sample and Loss Weight}

\begin{figure}[b]
\centering
\includegraphics[width=0.316\textwidth, height=0.3\textwidth]{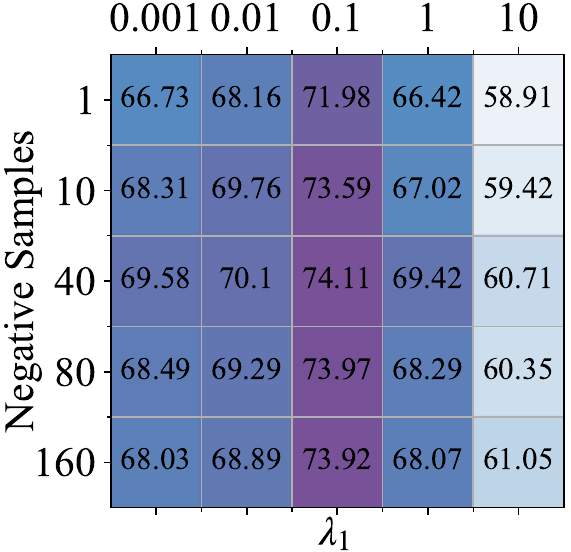}
\hspace{1cm}
\includegraphics[width=0.34\textwidth,height=0.3\textwidth]{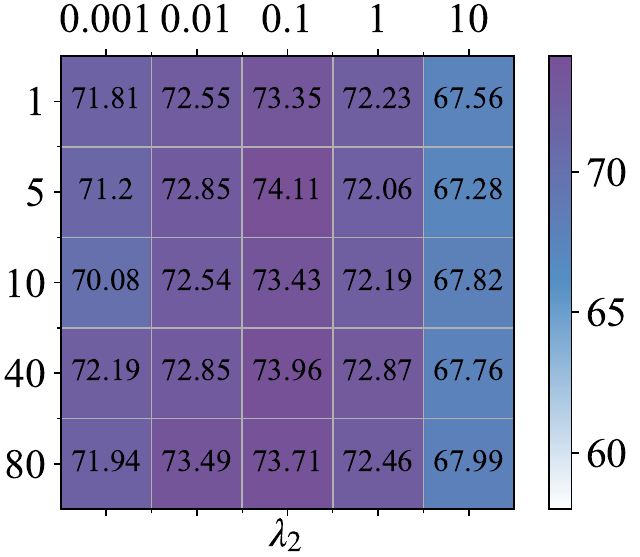}
\caption{ The performance of different numbers of negative samples and the loss weights in the risk minimization function for the contrastive learning component (left) and the disentangling component (right), respectively.}
\label{fig:loss_vs_negsample}
%\vskip -0.1in
%\vskip -0.1in
\end{figure}

The contrastive learning and disentangling components are both contrastive-based methods that require negative sampling. This section evaluates how the number of negative samples influences performance.
We also compare the influence of different loss weights of the two components.
We run our model with different negative sample numbers and loss weights for the two components, respectively.
From Figure \ref{fig:loss_vs_negsample}, we can see that a large loss weight, or a large number of negative samples does not necessarily result in a better performance.
Both components should be fine tuned to generate the best outcome.
Generally, a very large or small loss weight may make the multi-task training unbalanced, harming the final performance.
For the number of negative samples, a small number will make contrastive learning insufficient, while a large number may cause overfitting.

\subsection{Empirical Analysis of Time Complexity}
\label{sec:Emperical_timeComplexity}

\begin{table}[ht]
%\CenterFloatBoxes
\centering
\caption{ The overall time consumption of different models in one batch training.}
\label{tab:overall_time}
\small
\begin{tabular}{cc}
\toprule
Model & Time (ms) \\
\midrule
DCNv2 & 34.40  \\
AutoInt & 37.53 \\
SIGN & 40.41   \\
IEDR & 44.61 \\ 
\bottomrule
\end{tabular}
\end{table}

\begin{table}[ht]
\centering
\small
\caption{ The time consumption of critical procedures in IEDR in one batch training.}
\label{tab:component_time}
\begin{tabular}{cc}
\toprule
Procedure & Time (ms) \\
\midrule
Graph Learning (Feature Interaction Modeling) & 14.16 \\
  CICL  & 8.05   \\
  Disentangling (step 1)  & 0.16 \\
  Disentangling (step 2)  & 1.93  \\
  Optimization (step 1)     & 2.21  \\
  Optimization (step 2)     & 8.52  \\
\bottomrule
\end{tabular}
\end{table}

We summarize the overall time consumption of IEDR and several feature interaction-based baseline models in Table \ref{tab:overall_time}. The results are recorded by running the models for one batch (batch size 1024) on the Frappe dataset on a machine with CPU:12th Gen Intel(R) Core(TM) i9-12900K, RAM: 32GB, GPU: NVIDIA GeForce RTX 3090. 
We can see that our model's overall time consumption is only slightly higher than the other baselines. 
Next, we summarize the time cost of critical procedures in IEDR in Table \ref{tab:component_time}. The first four rows are model forwarding procedures, and the last two rows are model (alternative) optimizing procedures.
Table \ref{tab:component_time} shows the feature interaction modeling procedure takes most of the time, which is consistent with our analysis in Section \ref{sec:timeComplexityAnalysis}. 
CICL and disentangling forward procedures (rows 2-4) do not pose a large overhead since they reuse the feature interaction modeling results. 
Optimization (step 1) updates the parameters of the model's disentangling component ($q_1$ and $q_2$), which produces little overhead (2.21 ms) and is negligible in the whole procedure.

%{\bf Remark.} In addition to the above experiments, we further evaluate our model by studying how the hyperparameter settings influence the performance in Appendix \ref{appx:addi_exp}.

\subsection{Empirical Analysis of Falling Into Trivial Solutions}
\label{appx:exp_trivial}

\begin{figure}[t]
\centering
\includegraphics[width=0.25\textwidth]{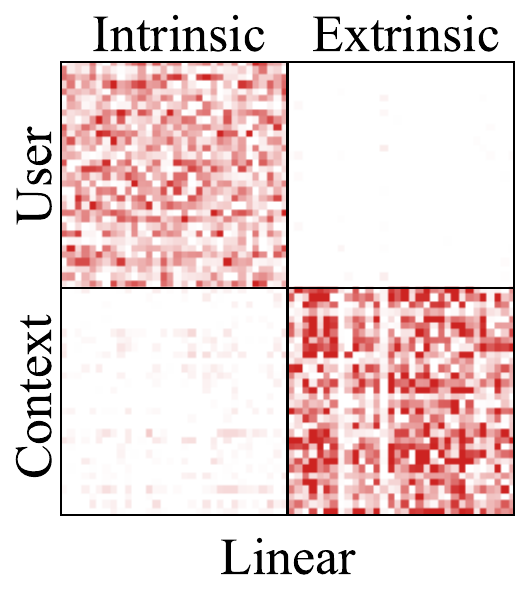}
\hspace{1cm}
\includegraphics[width=0.25\textwidth]{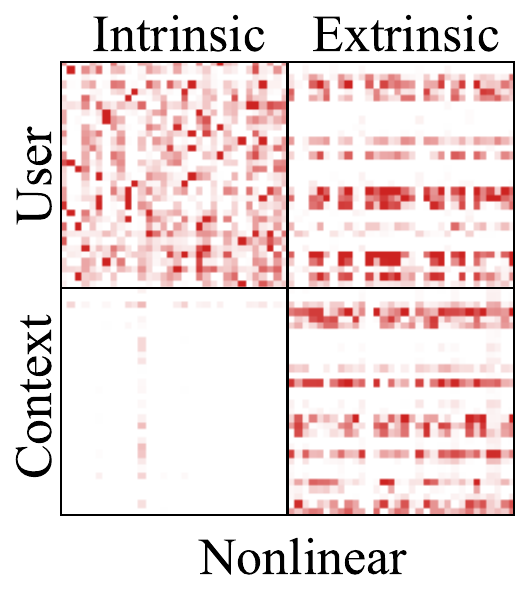}
\caption{ Visualization of $f_{ie}$ weights for the \textit{Linear} and \textit{Nonlinear} models.}
\label{fig:tri_weight}
\end{figure}

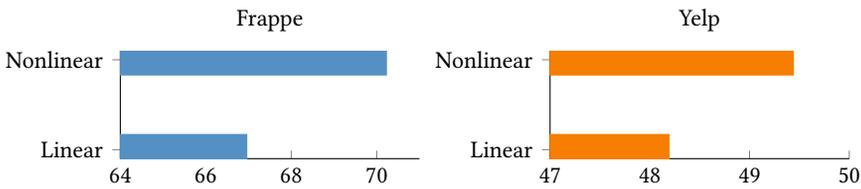
\begin{figure}[ht]
\centering
\begin{tikzpicture}
\begin{axis}[xbarstyle, height=3cm, width=0.40\textwidth, title={Frappe},bar width=4mm, symbolic y coords = {Linear, Nonlinear}, legend style={draw=none, at={(0.4,0.8)}, anchor=west, nodes={scale=0.9, transform shape}, legend image post style={scale=0.5}}, xmin=64,xmax=71]
\addplot [color=mycolor3, fill=mycolor3] coordinates { 
(66.959,Linear)
(70.226,Nonlinear)
};
\end{axis}
\end{tikzpicture}
\begin{tikzpicture}
\begin{axis}[xbarstyle, height=3cm, width=0.40\textwidth, title={Yelp},bar width=4mm, symbolic y coords = {Linear, Nonlinear}, legend style={draw=none, at={(0.4,0.8)}, anchor=west, nodes={scale=0.9, transform shape}, legend image post style={scale=0.5}}, xmin=47,xmax=50]
\addplot[color=mycolor1, fill=mycolor1] coordinates { 
(48.194,Linear)
(49.44,Nonlinear)
};
\end{axis}
\end{tikzpicture}
\caption{Comparing the performance of the \textit{Linear} and \textit{Nonlinear} models on different datasets.}
\label{fig:trivial_perfom}
\end{figure}

As discussed in Section \ref{appx:trivial_solution}, our model may fall into a trivial solution if $f_{ie}^{u}(\urep, \crep)$ is a linear mapping method. 
To evaluate how the trivial solution influences our model in learning the factors, we run our model with $f_{ie}$ being linear. 
Specifically, we concatenate $\urep$ and $\crep$ and feed them into an MLP without a hidden layer or activation (a linear mapping), making it easy to fall into the trivial solution. We call this variation \textit{Linear}. 
Then, we avoid this by simply adding a nonlinear activation function (ReLU) activation after the linear mapping. We call this variation \textit{Nonlinear}. 
Figure \ref{fig:tri_weight} shows the weight values of $f_{ie}$ of the two variations. The color shows the weights mapping from user/context representations to intrinsic/extrinsic representations. The darker the color, the larger the weight (the more information of user/context is mapped into intrinsic/extrinsic representations).
The figure shows that in the \textit{Linear} variation, user information is largely mapped into intrinsic representation (user-intrinsic block) but not extrinsic representation (user-extrinsic block). Context information is largely mapped into extrinsic representation (context-extrinsic block) but not intrinsic representation (context-intrinsic block). This means that the \textit{Linear} variation falls into the trivial solution. On the contrary, in the \textit{Nonlinear} variation, user information is mapped into extrinsic representation (user-extrinsic block), showing that the extrinsic representation contains both user and context information.
Figure \ref{fig:trivial_perfom} shows the performance of the two variations. We can see that the\textit{ Linear} model performs worse than the \textit{Nonlinear} model. It proves that learning intrinsic and extrinsic factors results in a better performance than simply mapping user and context information into two representations, respectively (the trivial solution).

\section{Conclusion}

%To improve the performance of recommender systems, we propose to accurately learn intrinsic and extrinsic factors from the interplay of various contexts, which more recently have been shown to be essential.
%To achieve that, we propose the intrinsic-extrinsic disentangled recommendation (IEDR) model. This generic model effectively learns intrinsic and extrinsic factors from various contexts for a more accurate recommendation. IEDR comprises a context-invariant contrastive learning component, and a mutual information minimization-based disentangling component to ensure the success of factor learning. Extensive experiments prove our model's ability to learn intrinsic and extrinsic factors for more accurate recommendation prediction.
%Following this work, we may explore learning more fine-grained intrinsic and extrinsic factors
%(e.g., multiple intrinsic factors) so that can capture nuanced users' interests and generalize our methods to broader applications, e.g., improving the diversity of recommendations. Also, we may explore how to disentangle intrinsic and extrinsic factors when context features are not available.

To enhance recommendation accuracy, we proposed IEDR, a novel framework that effectively differentiates and captures intrinsic and extrinsic factors from the interplay of various contexts. 
IEDR leverages a context-invariant contrastive learning component and a mutual information minimization-based disentangling component to capture consistent user preference and external motivation that may vary across contexts.
Extensive experiments on real-world datasets demonstrated IEDR's effectiveness in learning disentangled factors and significantly improving recommendation accuracy by up to 4\% in NDCG.
Following this work, we may explore learning more fine-grained intrinsic and extrinsic factors
(e.g., multiple intrinsic factors) so that can capture nuanced user interests and generalize our methods to broader applications, e.g., improving the diversity of recommendations. Also, we may explore how to disentangle intrinsic and extrinsic factors when context features are not available.

\section*{Acknowledgments}
This work was financially supported by the National Natural Science Foundation of China (Grant No. 62436003 and 62306333), ARC Discovery Project (Grant No. DP230102908 to Junhao Gan), and ARC Discovery Early Career Researcher Award (DECRA) (Grant No. DE220100680 to Sarah M. Erfani).

%This work is partly supported by the National Natural Science Foundation of China under grants 62436003. This work is partly supported by the National Natural Science Foundation of China under grants 62306333.
%In this work, Junhao Gan is in part supported by ARC Discovery Project DP230102908.

\bibliography{ref}
\bibliographystyle{ACM-Reference-Format}

\newpage
\clearpage
\setcounter{page}{1}
\appendix

\section{Proof of Theorem \ref{thm:theoryCICL}}
\label{appx:prooftheoryCICL}

\begin{proof}
Since the mutual information is not explicitly intractable, we approximate the right side of Equation (\ref{eq:CICLeqInfomaxmin}) with a lower bound (i.e., MINE \cite{belghazi2018mine}) and an upper bound (i.e., CLUB \cite{cheng2020club}) of mutual information, respectively. More formally,
\begin{equation}\label{eq:MINE}
\begin{split}
 \I(\Uin, \urep) \ge \I_{\textit{MINE}}(\Uin, \urep)  := \mathbb{E}_{p(\Uin,\urep)} \left[\log p(\Uin,\urep)\right]
 - \log \mathbb{E}_{p(\Uin)p(\urep)}\left[p(\Uin,\urep)\right],   \\
\end{split}
\end{equation}

\begin{equation}\label{eq:CLUB}
\begin{split}
  \I(\Uin, \crep) \!\le\! \I_{\textit{CLUB}}(\Uin, \crep):=  \mathbb{E}_{p(\Uin,\crep)}\left[\log p(\Uin|\crep)\right] 
- \mathbb{E}_{p(\Uin)p(\crep)}\left[\log p(\Uin|\crep)\right].  \\
\end{split}
\end{equation}
With the approximated terms above, proving Equation. (\ref{eq:CICLeqInfomaxmin}) turns to verify:
\begin{equation*}\label{eq:theoryCSSLapprox}
 \arg\min  \sum_{i=1}^{N}\mathcal{L}_{\text{CICL}}(u_i, c_i) \!=\! \arg\max \Bigl(\I_{\textit{MINE}}(\Uin, \urep) -  \I_{\textit{CLUB}}(\Uin, \crep)\Bigr).
\end{equation*}

% The Infomax between the intrinsic factor $\bm{v}_{I}$ and user representations $\urep$ using MINE \cite{belghazi2018mutual}:
% \begin{equation}
% \label{eq:Infomax_uv}
%     \max I_{MINE}(\bm{v}_I;\urep):= \max \mathbb{E}_{p(\bm{v}_I,\urep)}[\log p(\bm{v}_I,\urep)] - \log \mathbb{E}_{p(\bm{v}_I)p(\urep)}[p(\bm{v}_I,\urep)].
% \end{equation}
% The Infomin between the intrinsic factor $\bm{v}_{I}$ and context representations $\urep$ using CLUB \cite{cheng2020club}:
% \begin{equation}
% \label{eq:Infomin_cv}
%   \min I_{CLUB}(\bm{v}_I;\urep):= \min \mathbb{E}_{p(\bm{v}_I,\urep)}[\log p(\bm{v}_I|\urep)] - \mathbb{E}_{p(\bm{v}_I)p(\urep)}[\log p(\bm{v}_I|\urep)].
% \end{equation}

% Our contrastive learning objective can be formalized as:
% \begin{equation}
% \label{eq:cl}
%   \max CL(\bm{v}_I,\urep, \urep):= \max \mathbb{E}_{p(\bm{v}_I,\urep)p(\urep)}[\log p(\bm{v}_I|\urep,\urep)] - \log\mathbb{E}_{p(\bm{v}_I, \urep)p(\urep)}[ p(\bm{v}_I|\urep,\urep)].
% \end{equation}
% \begin{proposition}
% Optimizing the objective in Equation \eqref{eq:cl} is equivalent to simultaneously optimizing the objectives in Equation \eqref{eq:Infomax_uv} and Equation \eqref{eq:Infomin_cv}.
% \end{proposition}

By minimizing $\mathcal{L}_{\text{CICL}}$, we aim to make $(\Uin)_{ii}$ similar to $(\Uin)_{ij}$.
This procedure can be interpreted in probability as: increasing the probability of $f_{ie}^{u}(\urep_i, \crep_j)$ to predict $(\Uin)_{ii}$.
Therefore, maximizing the $\exp(\sim((\Uin)_{ii}, (\Uin)_{ij})/\tau)$ in Equation (\ref{eq:CICL_loss_CL}) is equivalent to maximizing $p((\Uin)_{ii}|\urep_i, \crep_j)$ ($\exp(\cdot)$ is monotone increasing so that does not influence the conclusion).
Similar to the above conclusion, minimizing $\exp(\sim((\Uin)_{ii}, (\Uin)_{\ell i})/\tau)$ is equivalent to minimizing $p((\Uin)_{ii}|\urep_\ell, \crep_i)$.
Therefore, we have 
\begin{equation*}
\begin{split}
    & - \sum_{i=1}^{N}\mathcal{L}_{\text{CICL}}(u_i, c_i) \\
    =& \sum_{i=1}^{N}\log \frac{\exp(\sim((\Uin)_{ii}, (\Uin)_{ij})/\tau)}{\sum_{u_\ell\in\mathcal{U}}\exp(\sim((\Uin)_{ii}, (\Uin)_{\ell i})/\tau)} \\
    =& \sum_{i=1}^{N}\log [\exp(\sim((\Uin)_{ii}, (\Uin)_{ij})/\tau)]
      - \sum_{i=1}^{N}\log [\sum_{u_\ell\in\mathcal{U}}\exp(\sim((\Uin)_{ii}, (\Uin)_{\ell i})/\tau)] \\
    =& \sum_{i=1}^{N}\log [p((\Uin)_{ii}|\urep_i, \crep_j)] - \sum_{i=1}^{N}\log [\sum_{u_\ell\in\mathcal{U}}p((\Uin)_{ii}|\urep_\ell, \crep_i)]. \\
\end{split}
\end{equation*}
    
Equation (\ref{eq:CICL_loss_CL}) only samples one context $c_j$ for each data point. However, during the training, all contexts in $\mathcal{C}$ are expected to be sampled. If we count all contexts, we have 
\begin{equation}\label{eq:CICL_loss_prob}
\begin{split}   
    & \sum_{i=1}^{N}\log [p((\Uin)_{ii}|\urep_i, \crep_j)] - \sum_{i=1}^{N}\log [\sum_{u_\ell\in\mathcal{U}}p((\Uin)_{ii}|\urep_\ell, \crep_i)] \\
    = & \sum_{i=1}^{N} \sum_{c_j\in\mathcal{C}}\log [p((\Uin)_{ii}|\urep_i, \crep_j)] \!-\! \sum_{i=1}^{N}\log [\sum_{u_\ell\in\mathcal{U}}p((\Uin)_{ii}|\urep_\ell, \crep_i)] \\
    = & \mathbb{E}_{p(\Uin,\urep)p(\crep)}[\log p(\Uin|\urep, \crep)] \!-\! \mathbb{E}_{p(\Uin, \crep)}\log\mathbb{E}_{p(\urep)}[ p(\Uin|\urep, \crep)]. \\
\end{split}
\end{equation}

Equation (\ref{eq:CICL_loss_prob}) is the probability form of the objective function of the context-invariant counteractive learning component (Equation (\ref{eq:CICL_loss_CL})). Equation (\ref{eq:CICL_loss_prob}) maximizes the likelihood $p(\Uin|\urep, \crep)$ given the joint distribution of users and intrinsic factors, with the marginal distribution of contexts. 
Meanwhile, it minimizes the likelihood $p(\Uin|\urep, \crep)$ given the joint distribution of contexts and intrinsic factors, with the marginal distribution of the user.\footnote{Note that only if $f_{ie}^{u}(\urep, \crep)$ is a many-to-one (or one-to-one) mapping then Equation (\ref{eq:CICL_loss_prob}) and Equation (\ref{eq:CICL_loss_CL}) will be equivalent. Otherwise, given a sample pair ($\urep, \crep$), $f_{ie}^{u}(\urep, \crep)$ may have different $\Uin$ outputs (i.e., one-to-many). In this situation, the first term of Equation (\ref{eq:CICL_loss_prob}) cannot guarantee that the same user with different context will have the same intrinsic factor (i.e., they may have various intrinsic factor representations while still meet the objective of the first term of Equation (\ref{eq:CICL_loss_prob})).
We use an MLP as $f_{ie}^{u}(\urep, \crep)$, which is a many-to-one mapping function. Therefore, we can ensure the equivalence between Equation (\ref{eq:CICL_loss_prob}) and Equation (\ref{eq:CICL_loss_CL}).} 

From Equation (\ref{eq:CICL_loss_prob}), we further have:

\begin{equation}
\label{eq:deriviation}
\begin{split}
& \mathbb{E}_{p(\Uin,\urep)p(\crep)}[\log p(\Uin|\urep, \crep)] - \mathbb{E}_{p(\Uin, \crep)}\log\mathbb{E}_{p(\urep)}[ p(\Uin|\urep, \crep)] \\
 \labelrel={myeq:equality} \,  & \mathbb{E}_{p(\Uin,\urep)p(\crep)}[\log p(\Uin|\urep,\crep)] - \mathbb{E}_{p(\Uin, \crep)p(\urep)}[\log p(\Uin|\urep,\crep)] \\
=\,  & \mathbb{E}_{p(\Uin,\urep)p(\crep)}[\log p(\Uin|\urep,\crep)]- \mathbb{E}_{p(\Uin, \crep)p(\urep)}[\log p(\Uin|\urep,\crep)]   + \left( \mathbb{E}_{p(\urep)}[\log p(\urep)] - \mathbb{E}_{p(\urep)}[\log p(\urep)] \right) \\
=\, & \mathbb{E}_{p(\Uin,\urep)p(\crep)}[\log p(\Uin|\urep,\crep)p(\urep)] - \mathbb{E}_{p(\Uin, \crep)p(\urep)}[\log p(\Uin|\urep,\crep)] - \mathbb{E}_{p(\urep)}[\log p(\urep)] \\
=\, & \mathbb{E}_{p(\Uin,\urep)p(\crep)}[\log p(\Uin,\urep|\crep)] - \mathbb{E}_{p(\Uin, \crep)p(\urep)}[\log p(\Uin|\urep,\crep)]  - \mathbb{E}_{p(\urep)}[\log p(\urep)] \\
=\, & \mathbb{E}_{p(\Uin,\urep)p(\crep)}[\log p(\Uin,\urep|\crep)] - \mathbb{E}_{p(\Uin, \crep)p(\urep)}[\log p(\Uin|\urep,\crep)] \\
& - \mathbb{E}_{p(\urep)}[\log p(\urep)]  +  \Bigl(\mathbb{E}_{p(\Uin)p(\urep)p(\crep)}[\log p(\Uin|\urep,\crep)] - \mathbb{E}_{p(\Uin)p(\urep)p(\crep)}[\log p(\Uin|\urep,\crep)] \Bigr) \\
= \, & \mathbb{E}_{p(\Uin,\urep)p(\crep)}[\log p(\Uin,\urep|\crep)] - \mathbb{E}_{p(\Uin)p(\urep)p(\crep)}[\log p(\Uin|\urep,\crep)] - \mathbb{E}_{p(\urep)}[\log p(\urep)] \\
&- \, \mathbb{E}_{p(\Uin, \crep)p(\urep)}[\log p(\Uin|\urep,\crep)] + \mathbb{E}_{p(\Uin)p(\urep)p(\crep)}[\log p(\Uin|\urep,\crep)] \\
=\, & \mathbb{E}_{p(\Uin,\urep)p(\crep)}[\log p(\Uin,\urep|\crep)] - \mathbb{E}_{p(\Uin)p(\urep)p(\crep)}[\log p(\Uin,\urep|\crep)] \\
& - \left(\mathbb{E}_{p(\Uin, \crep)p(\urep)}[\log p(\Uin|\urep,\crep)] - \mathbb{E}_{p(\Uin)p(\urep)p(\crep)}[\log p(\Uin|\urep,\crep)]\right) \\
=\, & \mathbb{E}_{p(\crep)}\left(\mathbb{E}_{p(\Uin,\urep)}[\log p(\Uin,\urep|\crep)] - \mathbb{E}_{p(\Uin)p(\urep)}[\log p(\Uin,\urep|\crep)]\right) \\
& -\, \mathbb{E}_{p(\urep)}\left(\mathbb{E}_{p(\Uin, \crep)}[\log p(\Uin|\urep,\crep)] - \mathbb{E}_{p(\Uin)p(\crep)}[\log p(\Uin|\urep,\crep)]\right).
\end{split}
\end{equation}

(\ref{myeq:equality}): In the second term, pushing the log inside the expectation does not change the minimizer. 
%in the case when the conditional distribution $p(\Uin|\urep, \crep)$ is uniform.
%Since $\log$ is monotonically increasing, adding or removing it does not influence the optimization direction.
%Maximizing the first term of Equation \eqref{eq:deriviation} has same effect of maximizing Equation \label{eq:Infomax_uv}.

Comparing Equation (\ref{eq:MINE}) and the first term of Equation (\ref{eq:deriviation}), they both act like classifiers whose objectives maximize the expected log-ratio of the joint distribution over the product of marginal distributions \cite{hjelm2018learning}.
Therefore, maximizing this term in Equation (\ref{eq:deriviation}) will have the same effect as maximizing Equation (\ref{eq:MINE}).
We can interpret the first term of Equation (\ref{eq:deriviation}) as maximizing the mutual information between users and the corresponding intrinsic factor, conditioned on a given context. 
Similarly, maximizing the negative of the second term of Equation (\ref{eq:deriviation}) will have the same effect of minimizing Equation (\ref{eq:CLUB}), which can be interpreted as minimizing the mutual information between contexts and the corresponding intrinsic factors, conditioned on a given user.

Therefore, we can conclude that:
% \begin{equation}
% \label{eq:final_result}
% \small
% \max CL(\bm{v}_I,\urep, \urep)\approx\max I_{MINE}(\bm{v}_I;\urep) - I_{CLUB}(\bm{v}_I;\urep).
% \end{equation}
\begin{equation*}
\begin{split}
   \arg\min  &\sum_{(u_i,v_i,c_i)\in\mathcal{D}}\mathcal{L}_{\text{CICL}}(u_i, c_i) \\
   &= \arg\max \I_{\textit{MINE}}(\Uin, \urep) -  \I_{\textit{CLUB}}(\Uin, \crep).
\end{split}
\end{equation*}
\end{proof}

\begin{table*}[ht]
\centering
\small
\caption{Implementation details of different variations on the recommendation prediction module. ``-'' represent the operation is the same as our original IEDR setting.}
\label{tab:implement_RP}
\begin{tabular}{lcc}
\toprule
Variation & \multicolumn{2}{c}{Recommendation Prediction Module} \\
 & \multicolumn{1}{c}{feature model\footnotemark[3]} & \multicolumn{1}{c}{$f_{ie}$ \footnotemark[4]} \\
\midrule
 IEDR & $\phi(\psi(\{MLP(\ufrep_{i}\odot\ufrep_{j})\}_{j\in u}))_{i\in u}\rightarrow\urep$ & $MLP(\urep\circ\crep)\rightarrow [\Uin, \Uex]$\\
\cmidrule(lr{0.5em}){1-3}
 AVG & $\psi(\ufrep_{i})_{i\in u}\to\urep$ & -\\
 MLP & $MLP(\psi(\ufrep_{i})_{i\in u})\to\urep$ & -\\
 BI & $\psi(\ufrep_{i}\odot\ufrep_{j})_{i,j\in u}\to\urep$ & -\\
\cmidrule(lr{0.5em}){1-3}
 Linear & - & $\bm{W}[\urep,\crep]\rightarrow [\Uin, \Uex]$\\
 Nonlinear & - & $\sigma(\bm{W}[\urep,\crep])\rightarrow [\Uin, \Uex]$\\
\cmidrule(lr{0.5em}){1-3}
 $\text{IEDR}_{sp}$ & - & $MLP_1(\urep)\rightarrow\Uin, MLP_2(\urep\circ\crep)\rightarrow\Uex$\\
\bottomrule
\end{tabular}
%\vskip -0.10in
%\footnotetext[0]{$^{\dagger}$Here we use user representation learning as an example. The item and context learning have the same structure. $\phi,\psi$ are both element-wise averaging functions and $\odot$ is the element-wise product.}
%\footnotetext[0]{$^{\ddagger}$Here we use user factor learning as an example. $\circ$ is a flexible operation to combine two vector, i.e., $\circ$ is element-wise product for the Frappe dataset, and element-wise summation for the Yelp dataset. $[\cdot,\cdot]$ is the concatenation operation. $\bm{W}$ is a linear transformation matrix, $\sigma$ is a ReLU activation.}
%\vskip -0.15in
\end{table*}

\begin{table*}[ht]
\centering
\small
\caption{Implementation details of different variations of the contrastive intrinsic-extrinsic disentanglement module. ``-'' represents the operation as the same as our original IEDR setting. $\times$ represents the variation that does not contain the component.}
\label{tab:implement_cied}
\begin{tabular}{p{0.1\textwidth}P{0.4\textwidth}P{0.4\textwidth}}
\toprule
Variation & \multicolumn{2}{c}{Contrastive Intrinsic-Extrinsic Disentangling Module} \\
 & Contrastive Learning Component\footnotemark[5] & Disentangling Component \\
\midrule
 IEDR & \textit{positive sample}: $\ufie(\urep_i, \crep_j)\rightarrow(\Uin)_{ij}$ \linebreak \textit{negative sample}: $\ufie(\urep_\ell, \crep_i)\rightarrow(\Uin)_{\ell i}$ $\ufie(\urep_\ell, \crep_j)\rightarrow(\Uin)_{\ell j}$ \linebreak $c_j=randChoice(\textit{NegGen1}, \textit{NegGen2})$ & $MLP_{\theta_1}(\Uin)\rightarrow(\Uex)^{\prime}$ ($q_{1}^{u}$) \linebreak $MLP_{\theta_2}(\Uex)\rightarrow(\Uin)^{\prime}$ ($q_{2}^{u}$)\\
\cmidrule(lr{0.5em}){1-3}
noDis & - & $\times$\\
noCL &$\times$& -\\
noCIED &$\times$& $\times$\\
\cmidrule(lr{0.5em}){1-3}
NegGen1 & $c_j$ is generated from \textit{NegGen1} & -\\
NegGen2 & $c_j$ is generated from \textit{NegGen2} & -\\
NegGen1\&2 & - & - \\
\cmidrule(lr{0.5em}){1-3}
vCLUB & - & $MLP_{\theta_1}(\Uin)\rightarrow(\Uex)^{\prime}$ ($q_{1}^{u}$)\\
BiDis & - & -\\
\cmidrule(lr{0.5em}){1-3}
 $\text{IEDR}_{sp}$ & \textit{positive sample}: $dropout((\Uin)_{i})\rightarrow(\Uin)^{p}$ \linebreak \textit{negative sample}: $dropout((\Uin)_\ell)\rightarrow(\Uin)^{n}$ & -\\
\bottomrule
\end{tabular}
%\vskip -0.10in
%\vskip -0.15in
\end{table*}

\footnotetext[3]{Here we use user representation learning as an example. The item and context learning have the same structure. $\phi,\psi$ are both element-wise averaging functions and $\odot$ is the element-wise product.}
\footnotetext[4]{Here we use user factor learning as an example. $\circ$ is a flexible operation to combine two vectors, i.e., $\circ$ is an element-wise product for the Frappe dataset, and an element-wise summation for the Yelp dataset. $[\cdot,\cdot]$ is the concatenation operation. $\bm{W}$ is a linear transformation matrix, $\sigma$ is a ReLU activation.}
\footnotetext[5]{For $\text{IEDR}_{sp}$, the positive samples $(\Uin)^{p}$ are generated through a dropout of the intrinsic representation of the user, and the negative samples $(\Uin)^{p}$ are generated through a dropout of intrinsic representations of random users.}

\section{Algorithm}
\label{appx:algorithm}

This section provides the training process of our IEDR model in Algorithm \ref{alg:trainingIEDR}. In each epoch, we use the batch stochastic gradient descent method.

\begin{algorithm}
\caption{Batch stochastic gradient descent training of IEDR.}
\label{alg:trainingIEDR}
\begin{algorithmic}[1]
\State \textbf{Input:} $\mathcal{D}=\{(u_i,v_i,c_i)\}_{i=1:N}$ with the corresponding true label $y_i$ for each data sample.
\State \textbf{Hyperparameters:} $B$: batch size; $L$: negative sample number for the context-invariant contrastive learning component; $L_{dis}$: negative sample number for the disentangling component.
\State \textbf{Parameters:} $\bm{\theta}_1^{u},\bm{\theta}_2^{u},\bm{\theta}_1^{v},\bm{\theta}_2^{v}$: parameters for $q_{1}^{u},q_{2}^{u},q_{2}^{v},q_{2}^{v}$, respectively; $\bm{\omega}$: parameters of IEDR except for $\bm{\theta}_1^{u},\bm{\theta}_2^{u},\bm{\theta}_1^{v},\bm{\theta}_2^{v}$.
\Function{ContrastiveLearning\_User}{$\{(\urep_i,\crep_i)\}_{i=1:B}$} 
    \For{$i=1,...,B$}
        \State $(\Uin)_{ii}\gets \ufie(\urep_i, \crep_i)$
        \State $ContextGen \gets RandomChoice(NegGen1, NegGen2)$
        \State $c_j\gets ContextGen(c_i)$
        \State $(\Uin)_{ij}\gets \ufie(\urep_i, \crep_j)$    \Comment{Generate positive samples.}
        \For{$\ell=1,...,L$} \Comment{Generate negative samples.}
            \State $u_{\ell_{1}}\gets randomChoice(\{u_i\}_{i=1:B})$, $(\Uin)_{\ell_{1}i} = \ufie(\urep_{\ell_{1}}, \crep_i) $
            \State $u_{\ell_{2}}\gets randomChoice(\{u_i\}_{i=1:B})$, $(\Uin)_{\ell_{2}j} = \ufie(\urep_{\ell_{2}}, \crep_j) $
        \EndFor
        \State $\mathcal{L}_{CICL}(u_i, c_i)\gets \text{Equation (\ref{eq:CICLeqInfomaxmin}) based on the above positive and negative samples}$
    \EndFor
    \State \Return $average(\{\mathcal{L}_{CICL}(u_i, c_i)\}_{i=1:B})$
\EndFunction
\Function{ContrastiveLearning\_Item}{$\{(\irep_i,\crep_i)\}_{i=1:B}$}
    \State Symmetric to \Call{ContrastiveLearning\_User}{}.
\EndFunction
\Function{Disentanglement\_User}{$\{(\urep_i,\crep_i)\}_{i=1:B}$}
        \For{$i=1,...,B$}
        \State $(\Uin)_{ii}, (\Uex)_{ii} \gets \ufie(\urep_i, \crep_i)$
        \State $(\Uex)_{ii}^{pred} \gets q_{\theta_1}((\Uin)_{ii}), (\Uin)_{ii}^{pred} \gets q_{\theta_2}((\Uex)_{ii})$ \Comment{Generate positive samples.}
        \State $a_{pos}^{\to} \gets MSE((\Uex)_{ii}, (\Uex)_{ii}^{pred}), a_{pos}^{\gets} \gets MSE((\Uin)_{ii}, (\Uin)_{ii}^{pred})$ 
        \State $a_{neg}^{\to}\gets 0, a_{neg}^{\gets}\gets 0$
        \For{$j=1,...,L_{dis}$}  \Comment{Generate negative samples.}
            \State $(\Uin)_r, (\Uex)_r \gets randomChoice\bigl(\{\bigl((\Uin)_{ii}, (\Uex)_{ii} \bigr)\}_{i=1:B}\bigr)$
            \State $(\Uex)_{r}^{pred} = q_{\theta_1}((\Uin)_{r}), (\Uin)_{r}^{pred} = q_{\theta_2}((\Uex)_{r})$
            \State $a_{neg}^{\to} \gets a_{neg}^{\to} + MSE((\Uex)_{ii}, (\Uex)_{r}^{pred})$
            \State $a_{neg}^{\gets} \gets a_{neg}^{\gets} + MSE((\Uin)_{ii}, (\Uin)_{r}^{pred})$
        \EndFor 
        \State $(\mathcal{L}_{bi\text{-}appr})_i \gets \frac{1}{2}(a_{pos}^{\to}+a_{pos}^{\gets})$
        \State $(\mathcal{L}_{Dis})_i \gets \frac{1}{2}(\frac{a_{neg}^{\to}+a_{neg}^{\gets}}{N_{dis}}-(a_{pos}^{\to}+a_{pos}^{\gets}))$
    \EndFor 
    \State \Return $average(\{(\mathcal{L}_{bi\text{-}appr})_i\}_{i=1:B}), average(\{(\mathcal{L}_{Dis})_i\}_{i=1:B})$ 
\EndFunction
\Function{Disentanglement\_Item}{$\{(\irep_i,\crep_i)\}_{i=1:B}$}
    \State Symmetric to \Call{Disentanglement\_User}{}.
\EndFunction
\State $ $
\algstore{myalg}
\end{algorithmic}
\end{algorithm}

\clearpage
\begin{algorithm}
  \ContinuedFloat
  \caption{Batch stochastic gradient descent training of IEDR (continued).}
  \begin{algorithmic}
      \algrestore{myalg}
      
\State $shuffle(\{(u_i, v_i, c_i)\}_{i=1:N})$
\For{\textit{each batch $\{(u_i, v_i,c_i)\}_{i=1:B}$}}
    \For{$i=1,...,B$} \Comment{Line 45-47 are the recommendation prediction module.}
        \State $\urep_i\gets f_u(u_i), \irep_i\gets f_v(v_i), \crep_i\gets f_c(c_i)$
        \State $(\Uin)_{ii}, (\Uex)_{ii} \gets \ufie(\urep_i, \crep_i)$, $(\Iin)_{ii}, (\Iex)_{ii} \gets \ifie(\irep_i, \crep_i)$
        \State $y^{\prime}_{i} \gets f_{pred}((\Uin)_{ii},(\Uex)_{ii}, (\Iin)_{ii}, (\Iex)_{ii})$
        \State $(\mathcal{L}_{RP})_i\gets CrossEntropy(y^{\prime}_{i}, y_i)$
    \EndFor
    \State $\mathcal{L}_{RP}\gets average(\{\mathcal{L}_{RP})_i\}_{i=1:B}$
    \State $\mathcal{L}_{CICL}^{u}\gets \Call{ContrastiveLearning\_User}{\{(\urep_i,\crep_i)\}_{i=1:B}}$
    \State $\mathcal{L}_{CICL}^{v}\gets \Call{ContrastiveLearning\_Item}{\{(\irep_i,\crep_i)\}_{i=1:B}}$
    \State $\mathcal{L}_{bi\text{-}appr}^{u},\mathcal{L}_{Dis}^{u} \gets \Call{Disentanglement\_User}{\{(\urep_i,\crep_i)\}_{i=1:B}}$
    \State $\mathcal{L}_{bi\text{-}appr}^{v},\mathcal{L}_{Dis}^{v} \gets \Call{Disentanglement\_Item}{\{(\irep_i,\crep_i)\}_{i=1:B}}$
    \State Freeze $\bm{\omega}$, update $\bm{\theta}_1^{u},\bm{\theta}_2^{u}, \bm{\theta}_1^{v},\bm{\theta}_2^{v} \text{ through minimizing } \mathcal{R}(\bm{\theta}_1^{u},\bm{\theta}_2^{u},\bm{\theta}_1^{v},\bm{\theta}_2^{v})$
    \Comment{Step 1}
    \State Freeze $\bm{\theta}_1^{u},\bm{\theta}_2^{u},\bm{\theta}_1^{v},\bm{\theta}_2^{v}$, update $\bm{\omega}$ through minimizing $\mathcal{R}(\bm{\omega})$
    \Comment{Step 2}
    \EndFor
  \end{algorithmic}
\end{algorithm}

\end{document}
\endinput
%%
%% End of file `sample-manuscript.tex'.